\documentclass[a4paper,12pt]{article}
\usepackage{amssymb}
\usepackage{amsmath}

\newcommand{\RSB}[2]{\makebox[#1]{$\left.\rule{0pt}{#2}\right]$}}
\newcommand{\LSB}[2]{\makebox[#1]{$\left[\rule{0pt}{#2}\right.$}}
\numberwithin{equation}{section}
\def\be{\begin{equation}}
\def\ee{\end{equation}}
\def\ba{\begin{array}}
\def\ea{\end{array}}

\newcommand{\bea}{\begin{eqnarray}}
\newcommand{\eea}{\end{eqnarray}}

\begin{document}

\begin{titlepage}

\begin{flushright}
CERN-PH-TH/2012-306
\end{flushright}

\vskip 0.5 cm
\begin{center}  {\Huge{\bf      Multi-Centered\\\vskip 0.3 cm First Order Formalism}}

\vskip 1.5 cm

{
{\bf Sergio Ferrara$^{1,3}$}, {\bf Alessio Marrani$^1$}, \\{\bf Andrey Shcherbakov$^3$}, {\bf Armen Yeranyan$^{2,3,4}$}}

\vskip 1.0 cm

$^1${\sl Physics Department, Theory Unit, CERN,\\CH 1211, Geneva 23, Switzerland\\
\texttt{sergio.ferrara@cern.ch}\\
\texttt{alessio.marrani@cern.ch}}\\

\vskip 0.5 cm

$^2${\sl Museo Storico della Fisica e Centro Studi e Ricerche \textquotedblleft\ Enrico Fermi''\\
Via Panisperna 89A, I-00184 Roma, Italy}

\vskip 0.5 cm

$^3${\sl INFN - Laboratori Nazionali di Frascati,\\Via Enrico Fermi
40, I-00044 Frascati, Italy\\
\texttt{ashcherb@lnf.infn.it}\\
\texttt{ayeran@lnf.infn.it}}\\

\vskip 0.5 cm

$^4${\sl Department of Physics, Yerevan State University\\
Alex Manoogian St. 1, Yerevan, 0025, Armenia}


 \end{center}

 \vskip 1.0 cm

\begin{abstract}

We propose a first order formalism for multi-centered black
holes with flat tree-dimensional base-space, within the $stu$ model of $N=2$,
$D=4$ ungauged Maxwell-Einstein supergravity. This provides a unified
description of first order flows of this universal sector of all models with
a symmetric scalar manifold which can be obtained by dimensional reduction
from five dimensions.

We develop a $D=3$ Cartesian formalism which suitably extends the definition
of central and matter charges, as well as of black hole effective potential
and first order ``fake'' superpotential, in order to deal
with not necessarily axisimmetric solutions, and thus with multi-centered
and/or (under-)rotating extremal black holes.

We derive general first order flow equations for composite non-BPS and
almost BPS classes, and we analyze some of their solutions, retrieving
various single-centered (static or under-rotating) and multi-centered known
systems.

As in the $t^{3}$ model, the almost BPS class turns out to split into two
general branches, and the well known almost BPS system is shown to be a
particular solution of the second branch.

 \end{abstract}
\vspace{24pt} \end{titlepage}



\section{\label{Intro}Introduction}

Multi-centered extremal black hole (BH) solutions in four-dimensional
supergravity theories have been widely investigated in recent years~\cite%
{Bossard:2011kz,Bossard-Octonionic,Goldstein-Katmadas,Bena,various,GLS-2,GP}.

Motivated by the issue of matching the true BPS spectrum and the spectrum of
spherically symmetric BHs in supergravity, in~\cite{Denef:2000nb} a
supersymmetric class of multi-centered BHs was introduced, whose BPS first
order flow equations were then solved and analyzed in~\cite{Denef-2},
revealing interesting features, such as fixed distance among the centers (in
presence of \textit{mutually non-local} electric-magnetic charge vectors).

However, only in the last two years a group-theoretical approach~\cite%
{Bossard:2011kz,Bossard-Octonionic}, based on nilpotent orbits and timelike
reduction to three dimensions, allowed for a systematic construction and
investigation of whole new classes of solutions, in which some or all BH
centers are non-supersymmetric (non-BPS).

An elegant approach to the flow dynamics of scalar fields in the background
of single-centered extremal BH solutions of Maxwell-Einstein theories of
(super)gravity, essentially based on the first order reformulation of the
scalar equations of motion, was introduced in~\cite{CD-1}, and then
developed in various works~\cite{FO-various,Bellucci:2008sv,Ceresole:2009iy}%
. The possibility to switch from second order to first order differential
equations of motion - \textit{without doubling their number} - has an
applicative relevance. Indeed, due to the interplay between auxiliary fields
and scalar charges, the first order formalism automatically discards
blowing-up solutions; furthermore, the integration of first order equations
is surely more manageable, and explicit forms of attractor flows can be more
easily determined.

The extension of such a formalism to non-supersymmetric multi-centered
configurations was started in~\cite{GP} by reducing the relevant action to a
sum of squares, however without yielding explicit expressions for the flow
equations and their corresponding governing functions. A consistent and
explicit determination of the most general first order flow equations for
non-BPS multi-centered and/or rotating BHs with flat three-dimensional
base-space was achieved in~\cite{Yeranyan-t^3}, within the simplest model of
$N=2$, $D=4$ Maxwell-Einstein ungauged supergravity with a cubic
prepotential, namely the so-called $t^{3}$ model, exhibiting only one vector
multiplet, and whose uplift to $D=5$ is \textit{``pure''} minimal
supergravity (see \textit{e.g.}~\cite{dWVVP}).

In the present investigation, we further develop the approach of~\cite%
{Yeranyan-t^3}, and determine the general first order flow equations for all
classes of multi-centered BH solutions with flat three-dimensional
base-space in the so-called $stu$ model~\cite{STU}; in this model, three
vector multiplets are coupled, in a \textit{triality-invariant} way, to the $%
N=2$ gravity multiplet, and the resulting completely factorized rank-$3$
symmetric special K\"{a}hler manifold $\left[ SU(1,1)/U(1)\right] ^{3}$ can
be considered a universal sector of all symmetric scalar manifolds of $D=4$,
$N\geqslant 2$-extended supergravity theories which admit a $D=5$ uplift.

A key result is the reformulation of second order equations of motion in a
manifestly $D=3$ Cartesian formalism, based on a timelike Lagrangian
reduction $D=4\rightarrow 3$ in a stationary BH background~\cite{BGM}, later
specialized for flat spatial slices. This formalism allows for a consistent
generalization of the BH effective potential~\cite{Ferrara:1997tw}, and of
its expression in terms of supersymmetry central charges and matter charges,
as well as in terms of a first order ``fake'' superpotential \cite{CD-1}, in
not necessarily axisymmetric contexts, which thus can include multi-centered
solutions.

Various under-rotating stationary (BPS and non-BPS) single-centered
solutions, as well as the known classes of \textit{BPS}~\cite{Denef:2000nb},
\textit{composite non-BPS}~\cite{Bossard:2011kz} and \textit{almost BPS}
\cite{Bossard:2011kz,Goldstein-Katmadas} multi-centered solutions (possibly
constrained in a suitable way) are retrieved as particular solutions of the
various first order systems which we propose. It should be stressed that,
analogously to~\cite{Yeranyan-t^3}, in this paper we focus on the general
formulation of first order systems of equations, leaving the determination
of new sets of solutions and the investigation of their physical properties
to future investigations.\medskip

The paper is organized as follows.

In Sec. \ref{2nd-order}, we start with the timelike reduction of
Maxwell-Einstein-scalar action in the background of stationary metrics (then
assumed to have flat spatial slices). As mentioned, this yields to a $D=3$
Cartesian formalism, which is especially suitable to handle multi-centered
and under-rotating single-centered solutions.

Our general approach to first order formalism is then exploited in Sec. \ref%
{1st-order} for the various classes of multi-centered solutions, namely the
\textit{BPS} (Sec. \ref{pSec-BPS}), \textit{almost BPS} (Sec. \ref%
{pSec-almost-BPS}) and \textit{composite non-BPS} (Sec. \ref%
{pSec-composite-nBPS}) classes, also retrieving and discussing known multi-
and single- centered solutions. In particular, as found in~\cite%
{Yeranyan-t^3} in the $t^{3}$ model, we find that the \textit{almost BPS}
class splits into two branches, which can essentially be related to the BPS
or non-BPS nature of the corresponding single-centered limits, as discussed
in Secs. \ref{Branch-I} and \ref{Branch-II}.

Finally, Sec. \ref{Conclusion} contains a summary and an outlook of results.

Some notation and useful formul\ae\ are given in App. \ref{notations}.

\section{\label{2nd-order}Second Order Equations of Motion}

Let us consider the $D=4$ Einstein-Maxwell action
\begin{equation}
S=\int d^{4}x\sqrt{\rule{0pt}{0.8em}-g}\left[ -\frac{1}{2}\,R+G_{a\bar{a}%
}g^{\mu \nu }\partial _{\mu }z^{a}\partial _{\nu }\overline{z}^{\bar{a}}+%
\frac{1}{4}\,\mu _{\Lambda \Sigma }F_{\mu \nu }^{\Lambda }F^{\Sigma \,\mu
\nu }+\frac{1}{4}\,\nu _{\Lambda \Sigma }F_{\mu \nu }^{\Lambda }\,^{\ast
}F^{\Sigma \,\mu \nu }\right] ,  \label{d4act}
\end{equation}%
which, depending on properties of the target space metric~$G_{a\overline{b}%
}(z,\overline{z})$ and of the coupling matrices~$\mu _{\Lambda \Sigma }(z,%
\overline{z})$ and~$\nu _{\Lambda \Sigma }(z,\overline{z})$, may be the
bosonic sector of some $N\geqslant 2$-extended, $D=4$ Maxwell-Einstein
supergravity. In the present investigation, we are interested in stationary
solutions to the Einstein-Maxwell equations of motion; we will thus perform
a timelike reduction to tree dimensions \textit{\`{a} la
Breitenlohner-Gibbons-Maison}~\cite{BGM}, giving rise to a $D=3$ Cartesian
formalism, which is not necessarily axisymmetric.

We start from a space-time metric satisfying the following \textit{Ansatz} ($%
i,j=1,2,3$):
\begin{equation}
ds^{2}=g_{\mu \nu }dx^{\mu }dx^{\nu }=e^{2U(x)}(dt+\omega
_{i}(x)dx^{i})^{2}-e^{-2U(x)}\gamma _{ij}(x)dx^{i}dx^{j}.
\label{stationaryMetr}
\end{equation}%
In this background, the Abelian vector potential~$A_{\mu }^{\Lambda }$
(defining the two-form field strength\footnote{%
Throughout the paper, the (anti)symmetrization is defined with a $1/n!$
normalization if applied to~$n$ indices.} $F_{\mu \nu }^{\Lambda }=2\partial
_{\lbrack \mu }A_{\nu ]}^{\Lambda }$) splits into $D=3$ vector potentials~$%
a_{i}^{\Lambda }$ and Kaluza-Klein scalars~$b^{\Lambda }$:
\begin{equation*}
A_{\mu }^{\Lambda }dx^{\mu }=b^{\Lambda }(dt+\omega
_{i}dx^{i})+a_{i}^{\Lambda }dx^{i}.
\end{equation*}%
%
%
%
%
%
By using the $D=3$ vectors~$a_{i}^{\Lambda }$ and~$\omega _{i}$, one can
construct the corresponding field strengths
\begin{equation*}
f_{ij}^{\Lambda }=2\partial _{\lbrack i}a_{j]}^{\Lambda },\qquad
W_{ij}=2\partial _{\lbrack i}\omega _{j]},
\end{equation*}%
%
%
%
%
%
respectively expressing the magnetic field and the rotation in tree
dimensions.

On stationary background~(\ref{stationaryMetr}), the action~(\ref{d4act}) is
equivalent to the following $D=3$ one:
\begin{equation}
\begin{array}{l}
\displaystyle S=\int d^{\,3}x\sqrt{\rule{0pt}{0.8em}\gamma }\left[ -\frac{1}{%
2}\,R^{(3)}+G_{a\bar{a}}\gamma ^{ij}\partial _{i}z^{a}\partial _{j}\overline{%
z}^{\bar{a}}+\gamma ^{ij}\partial _{i}U\partial _{j}U-\frac{1}{8}%
\,e^{4U}W_{ij}W^{ij}\right. \\
\displaystyle\phantom{S=\int d^{\,3}x} +\frac{1}{2}\,e^{-2U}\mu _{\Lambda
\Sigma }\gamma ^{ij}\partial _{i}b^{\Lambda }\partial _{j}b^{\Sigma }-\frac{1%
}{4}\,e^{2U}\mu _{\Lambda \Sigma }\left( f_{ij}^{\Lambda }+b^{\Lambda
}W_{ij}\right) \left( f^{\Sigma \,ij}+b^{\Sigma }W^{ij}\right) \\
\displaystyle\phantom{S=\int d^{\,3}x} \left. -\nu _{\Lambda \Sigma
}\,\partial _{i}b^{\Lambda }\left( ^{\ast }f^{\Sigma \,i}+b^{\Sigma }~^{\ast
}W^{i}\right) \right] ,%
\end{array}
\label{d3act}
\end{equation}%
where the $D=3$ Hodge operator ``$\ast $'' is defined with respect to the
3-dimensional metric~$\gamma _{ij}$:
\begin{equation*}
^{\ast }A^{i}\equiv \frac{1}{2\sqrt{\rule{0pt}{0.8em}\gamma }}\,\varepsilon
^{ijk}A_{jk}.
\end{equation*}%
%
%
%
%
%
The $D=3$ two-forms~$f_{ij}^{\Lambda }$ and~$W_{ij}$ can be dualized into
scalars~$b_{\Lambda }$ and~$\psi $ by adding the following terms to the
action~(\ref{d3act})
\begin{equation*}
\int d^{\,3}x\left[ \frac{1}{2}\,b_{\Lambda }\epsilon ^{ijk}\partial
_{i}f_{jk}^{\Lambda }+\frac{1}{2}\,\psi \,\epsilon ^{ijk}\partial _{i}W_{jk}%
\right] ,  \label{dualization}
\end{equation*}%
%
%
%
%
%
thus allowing for the two-forms to be expressed in terms of their duals:
\begin{equation}
\begin{array}{ll}
\displaystyle f^{\Lambda \,ij}+b^{\Lambda }W^{ij}=-\frac{1}{\sqrt{\rule%
{0pt}{0.8em}\gamma }}\,e^{-2U}\,\mu ^{\Lambda \Sigma }\varepsilon
^{ijk}\left( \partial _{k}b_{\Sigma }+\nu _{\Sigma \Sigma ^{\prime
}}\partial _{k}b^{\Sigma ^{\prime }}\right) ; & \displaystyle\qquad %
\mbox{(a)} \\
\displaystyle W^{ij}=-\frac{2}{\sqrt{\rule{0pt}{0.8em}\gamma }}%
\,e^{-4U}\varepsilon ^{ijk}\left( \partial _{k}\psi -b^{\Lambda }\partial
_{k}b_{\Lambda }\right) , & \displaystyle\qquad \mbox{(b)}%
\end{array}
\label{fW}
\end{equation}%
as well as the action~(\ref{d3act})to be recast as:
\begin{equation}
\begin{array}{l}
\displaystyle S=\int d^{\,3}x\sqrt{\rule{0pt}{0.8em}\gamma }\left[ -\frac{1}{%
2}\,R^{(3)}+G_{a\bar{a}}\gamma ^{ij}\partial _{i}z^{a}\partial _{j}\overline{%
z}^{\bar{a}}+\gamma ^{ij}\partial _{i}U\partial _{j}U+\frac{1}{2}%
\,e^{-2U}\mu _{\Lambda \Sigma }\gamma ^{ij}\partial _{i}b^{\Lambda }\partial
_{j}b^{\Sigma }\right. \\
\displaystyle\phantom{S=\int d^{\,3}x} +\,\frac{1}{2}\,e^{-2U}\mu ^{\Lambda
\Sigma }\gamma ^{ij}\left( \partial _{i}b_{\Lambda }+\nu _{\Lambda \Lambda
^{\prime }}\partial _{i}b^{\Lambda ^{\prime }}\right) \left( \partial
_{j}b_{\Sigma }+\nu _{\Sigma \Sigma ^{\prime }}\partial _{j}b^{\Sigma
^{\prime }}\right) \\
\displaystyle\phantom{S=\int d^{\,3}x}\left. +\, e^{-4U}\gamma ^{ij}\left(
\partial _{i}\psi -b^{\Lambda }\partial _{i}b_{\Lambda }\right) \left(
\partial _{j}\psi -b^{\Sigma }\partial _{j}b_{\Sigma }\right) \right] .%
\end{array}
\label{d3act2}
\end{equation}

By grouping the Kaluza-Klein scalars~$b^{\Lambda }$ and the scalars~$%
b_{\Lambda }$ (dual to~$f_{ij}^{\Lambda }$) into a symplectic vector~$%
b^{\alpha }$ (with $\alpha $ running over contravariant and covariant
symplectic index $\Lambda $) and introducing a scalar~$\tilde{\psi}$
\begin{equation}
b^{\alpha }\equiv \left(
\begin{array}{c}
b^{\Lambda } \\
b_{\Lambda }%
\end{array}%
\right) ,\qquad \tilde{\psi}\equiv 2\psi -b^{\Lambda }b_{\Lambda },
\label{b}
\end{equation}%
the action~(\ref{d3act2}) can be rewritten in more compact form:
\begin{equation}
\begin{array}{l}
\displaystyle S=\int d^{\,3}x\sqrt{\rule{0pt}{0.8em}\gamma }\left[ -\frac{1}{%
2}\,R^{(3)}+G_{a\bar{a}}\gamma ^{ij}\partial _{i}z^{a}\partial _{j}\overline{%
z}^{\bar{a}}+\gamma ^{ij}\partial _{i}U\partial _{j}U+\frac{1}{2}%
\,e^{-2U}\gamma ^{ij}\partial _{i}b^{\alpha }M_{\alpha \beta }\partial
_{j}b^{\beta }\right. \\
\displaystyle \phantom{S=\int d^{\,3}x} \left. +\frac{1}{4}\,e^{-4U}\gamma
^{ij}\left( \partial _{i}\tilde{\psi}+\langle b,\partial _{i}b\rangle
\right) \left( \partial _{j}\tilde{\psi}+\langle b,\partial _{j}b\rangle
\right) \right] .%
\end{array}
\label{d3act3}
\end{equation}%
For the definitions of the symplectic symmetric matrix~$M_{\alpha \beta }$
and the skew-symmetric symplectic product~$\langle \cdot ,\cdot \rangle $,
see App.~\ref{notations}. Then, by introducing a symplectic vector~$%
f_{ij}^{\alpha }=\left( f_{ij}^{\Lambda },f_{\Lambda \,ij}\right) ^{%
{\scriptsize \intercal }}$ constructed from the field strengths, Eqs.~(\ref%
{fW}) enjoy a manifest symplectic covariance\footnote{%
Eq.~(\ref{fW}a) is just the upper half of Eq.~(\ref{fsymplCov}a).}:
\begin{equation}
\begin{array}{ll}
\displaystyle f_{ij}^{\alpha }=\sqrt{\rule{0pt}{0.8em}\gamma }\,\epsilon
_{ijk}\,\gamma ^{kk^{\prime }}e^{-2U}M^{\alpha \beta }\Omega _{\beta \gamma
}\partial _{k^{\prime }}b^{\gamma }-b^{\alpha }W_{ij}; & \displaystyle\qquad %
\mbox{(a)} \\
\displaystyle W_{ij}=-\sqrt{\rule{0pt}{0.8em}\gamma }\,e^{-4U}\epsilon
_{ijk}\gamma ^{kl}\left( \partial _{l}\tilde{\psi}+\langle b,\partial
_{l}b\rangle \right) . & \displaystyle\qquad \mbox{(b)}%
\end{array}
\label{fsymplCov}
\end{equation}
The second order equations of motion pertaining to the action~(\ref{d3act3})
consist of the Einstein equations
\begin{equation}
\begin{array}{l}
\displaystyle \frac{1}{2}R_{ij}^{(3)}=G_{a\overline{a}}\partial
_{(i}z^{a}\partial _{j)}\overline{z}^{\overline{a}}+\partial _{i}U\partial
_{j}U+\frac{1}{2}\,e^{-2U}\partial _{i}b^{\alpha }M_{\alpha \beta }\partial
_{j}b^{\beta } \\
\displaystyle\phantom{\frac{1}{2}R_{ij}^{(3)}=G_{a\overline{a}}\partial
_{(i}z^{a}\partial _{j)}\overline{z}^{\overline{a}}}+ \frac{1}{4}%
\,e^{-4U}\left( \partial _{i}\tilde{\psi}+\langle b,\partial _{i}b\rangle
\right) \left( \partial _{j}\tilde{\psi}+\langle b,\partial _{j}b\rangle
\right) , \label{einst}%
\end{array}%
\end{equation}%
and of the following ones:
\begin{equation}
\begin{array}{l}
\begin{array}{c}
\partial _{i}\left[ \sqrt{\rule{0pt}{0.8em}\gamma }\gamma ^{ij}e^{-4U}\left(
\partial _{j}\tilde{\psi}+\langle b,\partial _{j}b\rangle \right) \right] =0,
\\
\end{array}
\\[0.5em]
\begin{array}{l}
\partial _{i}\left[ \sqrt{\rule{0pt}{0.8em}\gamma }\gamma
^{ij}e^{-2U}M_{\alpha \beta }\partial _{j}b^{\beta }\right] =\partial _{i}%
\left[ \sqrt{\rule{0pt}{0.8em}\gamma }\,\gamma ^{ij}e^{-4U}\Omega _{\alpha
\beta }b^{\beta }\,\left( \partial _{j}\tilde{\psi}+\langle b,\partial
_{j}b\rangle \right) \right] , \\
\end{array}
\\[0.5em]
\begin{array}{l}
\displaystyle \frac{1}{\sqrt{\rule{0pt}{0.8em}\gamma }}\,\partial _{i}\left[
\sqrt{\rule{0pt}{0.8em}\gamma }\gamma ^{ij}\partial _{j}U\right] =-\frac{1}{2%
}\,e^{-2U}\gamma ^{ij}\partial _{i}b^{\alpha }M_{\alpha \beta }\partial
_{j}b^{\beta } \\
\phantom{\frac{1}{\sqrt{\rule{0pt}{0.8em}\gamma }}\,\partial _{i}\left[
\sqrt{\rule{0pt}{0.8em}\gamma }\gamma ^{ij}\partial _{j}U\right] =} -\frac{1%
}{2}\,e^{-4U}\gamma ^{ij}\left( \partial _{i}\tilde{\psi}+\langle b,\partial
_{i}b\rangle \right) \left( \partial _{j}\tilde{\psi}+\langle b,\partial
_{j}b\rangle \right) , \\
\end{array}
\\
\begin{array}{l}
\displaystyle\frac{1}{\sqrt{\rule{0pt}{0.8em}\gamma }}\,\partial _{i}\left[
\sqrt{\rule{0pt}{0.8em}\gamma }\gamma ^{ij}G_{a\overline{a}}\partial _{j}%
\overline{{z}}^{\overline{a}}\right] =\frac{\partial G_{b\bar{b}}}{\partial
z^{a}}\gamma ^{ij}\partial _{i}z^{b}\partial _{j}\overline{{z}}^{\overline{b}%
}+\frac{1}{2}\,e^{-2U}\gamma ^{ij}\partial _{i}b^{\alpha }\frac{\partial
M_{\alpha \beta }}{\partial z^{a}}\,\partial _{j}b^{\beta }. \\
\end{array}%
\end{array}
\label{eom}
\end{equation}

Within the class of the stationary metrics~(\ref{stationaryMetr}), we
further select those with $D=3$ Euclidean metric~$\gamma _{ij}=\delta _{ij}$%
:
\begin{equation}
ds^{2}=e^{2U}(dt+\omega _{i}dx^{i})^{2}-e^{-2U}dx^{i}dx^{i}.
\label{stationaryMetrFlat}
\end{equation}%
Thus, we will henceforth restrict \ to consider $D=4$ under-rotating BHs,
with \textit{flat} $D=3$ base space. When replacing~(\ref{stationaryMetr})
with~(\ref{stationaryMetrFlat}), Einstein equations~(\ref{einst}) become a
first order constraint:
\begin{eqnarray}
G_{a\bar{a}}\partial _{(i}z^{a}\partial _{j)}\overline{z}^{\bar{a}}+\partial
_{i}U\partial _{j}U+\frac{1}{2}\,e^{-2U}\partial _{i}b^{{\scriptsize %
\intercal }}M\partial _{j}b+\frac{1}{4}\,e^{-4U}\left( \partial _{i}\tilde{%
\psi}+\langle b,\partial _{i}b\rangle \right) \left( \partial _{j}\tilde{\psi%
}+\langle b,\partial _{j}b\rangle \right) &=&0,  \notag \\
&&  \label{EinstConstr}
\end{eqnarray}%
and the equations of motion~(\ref{eom}) get slightly simplified:
\begin{equation}
\begin{array}{ll}
\displaystyle\partial _{i}\left[ e^{-4U}\left( \partial _{i}\tilde{\psi}%
+\langle b,\partial _{i}b\rangle \right) \right] =0, & \quad \mbox{(a)} \\%
[0.5em]
\displaystyle\partial _{i}\left[ e^{-2U}M_{\alpha \beta }\partial
_{i}b^{\beta }\right] =\partial _{i}\left[ e^{-4U}\Omega _{\alpha \beta
}b^{\beta }\,\left( \partial _{i}\tilde{\psi}+\langle b,\partial
_{i}b\rangle \right) \right] , & \quad \mbox{(b)} \\[0.5em]
\displaystyle\partial _{i}\partial _{i}U=-\frac{1}{2}\,e^{-2U}\partial
_{i}b^{{\scriptsize \intercal }}M\partial _{i}b-\frac{1}{2}\,e^{-4U}\left(
\partial _{i}\tilde{\psi}+\langle b,\partial _{i}b\rangle \right) ^{2}, &
\quad \mbox{(c)} \\
\displaystyle\partial _{i}\left[ G_{a\bar{a}}\partial _{i}\overline{z}^{\bar{%
a}}\right] =\frac{\partial G_{b\bar{b}}}{\partial z^{a}}\,\partial
_{i}z^{b}\partial _{i}\overline{z}^{\bar{b}}+\frac{1}{2}\,e^{-2U}\partial
_{i}b^{{\scriptsize \intercal }}\frac{\partial M}{\partial z^{a}}\,\partial
_{i}b. & \quad \mbox{(d)}%
\end{array}
\label{eom2}
\end{equation}
As a notation, let us introduce a $D=3$ vector~$\chi _{i}$, which, by virtue
of~(\ref{eom2}a), is divergenceless:
\begin{equation}
\chi _{i}\equiv e^{-4U}\left( \partial _{i}\tilde{\psi}+\langle b,\partial
_{i}b\rangle \right) ,\qquad \partial _{i}\chi _{i}=0.  \label{divergenless}
\end{equation}%
Maxwell equations~(\ref{eom2}b) can immediately be integrated, yielding:
\begin{equation}
\partial _{i}b^{\alpha }=e^{2U}M^{\alpha \beta }\Omega _{\beta \gamma }\hat{H%
}_{i}^{\gamma },\qquad \hat{H}_{i}^{\alpha }\equiv H_{i}^{\alpha }+b^{\alpha
}\chi _{i},\qquad \partial _{i}H_{i}^{\alpha }=0,  \label{maxwellIntegrated}
\end{equation}%
where another $D=3$ divergenceless vector~$H_{i}^{\alpha }$ has been
defined. The electromagnetic two-form field strength~$f_{ij}^{\alpha }$ and
the two-form~$W_{ij}$ corresponding to rotation are then just Hodge duals of
the vectors~$\chi _{i}$ and~$\hat{H}_{i}^{\alpha }$ defined in~(\ref%
{divergenless})-(\ref{maxwellIntegrated}):
\begin{equation*}
W_{ij}=-\varepsilon _{ijk}\chi _{k},\qquad f_{ij}^{\alpha }=-\varepsilon
_{ijk}\hat{H}_{k}^{\alpha }-b^{\alpha }W_{ij}=-\varepsilon
_{ijk}H_{k}^{\alpha }.
\end{equation*}%
%
%
%
%
%
This allows one to rewrite the Einstein equations~(\ref{EinstConstr}) as
well as the rest of the equations of motion~(\ref{eom2}) as follows:
\begin{eqnarray}
&&%
\begin{array}{l}
G_{a\bar{a}}\partial _{(i}z^{a}\partial _{j)}\overline{z}^{\bar{a}}+\partial
_{i}U\partial _{j}U-\hat{V}_{ij}e^{2U}+\frac{1}{4}\,e^{4U}\chi _{i}\chi
_{j}=0,%
\end{array}
\label{Einst2} \\
&&  \notag \\
&&%
\begin{array}{l}
\partial _{i}\partial _{i}U=e^{2U}\hat{V}_{ii}-\frac{1}{2}\,e^{4U}\chi
_{i}^{2}, \\[0.5em]
\partial _{i}\strut \left( G_{a\bar{a}}\partial _{i}\overline{z}^{\bar{a}%
}\right) =\partial _{a}G_{b\bar{b}}\,\partial _{i}z^{b}\partial _{i}%
\overline{z}^{\bar{b}}+e^{2U}\partial _{a}\hat{V}_{ii}.%
\end{array}
\label{eom3}
\end{eqnarray}

In~(\ref{Einst2})-(\ref{eom3}) the $D=3$ Cartesian tensor
\begin{equation}
\hat{V}_{ij}\equiv -\frac{1}{2}\hat{H}_{i}^{{\scriptsize \intercal }}M\hat{H}%
_{j}  \label{V-Cart}
\end{equation}%
has been introduced; it will be referred to as \textit{black hole potential}%
. In fact, it generalizes, for the whole class of metrics~(\ref%
{stationaryMetrFlat}), the well known BH potential~$V_{BH}$.

This latter was introduced in~\cite{Ferrara:1997tw} for the case of a static
and spherically symmetric BH. In our approach, this case corresponds to a
vanishing~$\chi _{i}$ and to the choice of the divergenceless vector~$%
H_{i}^{\alpha }$ to be the gradient of an harmonic function:
\begin{equation}
\chi _{i}=0,\qquad H_{i}^{\alpha }=\partial _{i}H^{\alpha }.  \label{1-ctr}
\end{equation}%
In the case of spherical symmetry, $H^{\alpha }$ has a single pole where the
black hole horizon resides (\textit{i.e.} in the origin):
\begin{equation}
H^{\alpha }(x,y,z)=h^{\alpha }+P^{\alpha }\tau ,\qquad h^{\alpha }=%
\mbox{const},\qquad \tau =\frac{1}{\sqrt{x^{2}+y^{2}+z^{2}}}.
\label{harmonicH}
\end{equation}%
The constants~$P^{\alpha }$ fit into the symplectic vector of
electromagnetic charges~($p^{\Lambda }$,$q_{\Lambda }$) of the BH itself.
Within these assumptions, the tensor~$\hat{V}_{ij}$ relates to~$V_{BH}$ as
follows:
\begin{equation}
\hat{V}_{ij}=\partial _{i}\tau \partial _{j}\tau V_{BH},\qquad V_{BH}=-\frac{%
1}{2}P^{{\scriptsize \intercal }}MP.  \label{1-ctr-2}
\end{equation}%
Moreover, Maxwell Eqs.~(\ref{maxwellIntegrated}) decouple completely from~(%
\ref{Einst2}) and~(\ref{eom3}), which get exactly the form presented in~\cite%
{Ferrara:1997tw}. Therein, it was also shown that in the case in which the
action~(\ref{d4act}) describes the bosonic sector of $N=2$, $D=4$
Maxwell-Einstein supergravity, the BH potential~$V_{BH}$ acquires a nice
geometrical interpretation in terms of \textit{special K\"{a}hler geometry}
(see \textit{e.g.}~\cite{CDF-rev} for a review and a list of Refs.):
\begin{equation}
V_{BH}=Z\bar{Z}+G^{a\bar{a}}D_{a}Z\bar{D}_{\bar{a}}\bar{Z},  \label{defVBH}
\end{equation}%
where~$Z$ is the $N=2$ central charge function and~$D_{a}Z$ (matter charges)
denotes its K\"{a}hler-covariant derivative:
\begin{equation}
Z=\langle P,V\rangle ,\qquad D_{a}Z=\langle P,D_{a}V\rangle ,  \label{std-1}
\end{equation}%
defined in terms of the symplectic sections~$V^{\alpha }$ of the flat
symplectic bundle of special geometry~\cite{Strominger-SKG,N=2-Big}.

Interestingly, the Cartesian $D=3$ tensor BH potential~$\hat{V}_{ij}$~(\ref%
{V-Cart}) can be given a geometrical interpretation in the spirit of Eq.~(%
\ref{defVBH}). Indeed, if one constructs a $D=3$ \textit{Cartesian
generalization} of the central charge and of its covariant derivatives as
\begin{equation}
\hat{Z}_{i}\equiv \langle \hat{H}_{i},V\rangle ,\qquad \hat{Z}_{ai}\equiv
\langle \hat{H}_{i},D_{a}V\rangle ,  \label{Zhat}
\end{equation}%
then it can be computed that
\begin{equation}
\hat{V}_{ij}=\hat{Z}_{(i}\hat{\overline{Z}}_{j)}+G^{a\bar{a}}\hat{Z}_{a(i}%
\hat{\overline{Z}}_{\bar{a}j)},  \label{Ress-1}
\end{equation}%
which can be regarded as the generalization of~(\ref{defVBH}) to generic,
not necessarily axisymmetric, and thus possibly multi-centered, stationary
solutions (at least those with flat three-dimensional spatial slices,
\textit{cfr.}~(\ref{stationaryMetrFlat})).

For later convenience, let us also here define the non-rotating limit ($\chi
_{i}=0$) of $\hat{Z}_{i}$ and $\hat{Z}_{ai}$ defined in~(\ref{Zhat}):
\begin{equation}
Z_{i}=\langle H_{i},V\rangle ,\qquad Z_{ai}=\langle H_{i},D_{a}V\rangle .
\label{Z}
\end{equation}%
As it will become evident from the subsequent treatment, both Cartesian $D=3$
generalizations~(\ref{Zhat}) and~(\ref{Z}) of the $N=2$ central charge $Z$
and of its covariant derivatives $D_{a}Z$ will play an important role in the
construction of first order equations of motion for multi-centered BHs.

\section{\label{1st-order}First Order Equation of Motion}

The first order equations of motion arise from the following \textit{Ansatz}%
:
\begin{equation}
\partial _{i}U=e^{U}W_{i},\quad \partial _{i}z^{a}=e^{U}\Pi _{i}^{a},\quad
\chi _{i}=2e^{-U}\ell _{i},  \label{FO-1}
\end{equation}%
where~$W_{i}$,~$\Pi _{i}^{a}$ and~$\ell _{i}$ are functions of the scalar
fields and, eventually, of some auxiliary ones. The choice of the \textit{%
Ansatz}~(\ref{FO-1}) is motivated by the fact that for single- and multi-
centered BPS, as well for single-centered non-BPS BHs, the first order
equations are of the form~(\ref{FO-1}).

In the following treatment, we will conveniently consider special K\"{a}hler
geometry tensors whose indices are \textquotedblleft
flattened\textquotedblright\ as usual:
\begin{equation}
V_{{\underline{a}}}\equiv E_{{\underline{a}}}{}^{a}D_{a}V,\quad \hat{Z}_{{%
\underline{a}}\,i}\equiv E_{{\underline{a}}}{}^{a}\hat{Z}_{a\,i}\qquad %
\mbox{etc.,}  \label{flattening}
\end{equation}%
by using the \textit{Vielbein} of the scalar manifold (for further details,
see App.~\ref{notations}). By means of~(\ref{flattening}), the first order
equations~(\ref{FO-1}) can be recast as follows:
\begin{equation}
\partial _{i}U=e^{U}W_{i},\qquad \partial _{i}z^{a}=e^{U}\Pi _{i}^{{%
\underline{a}}}\,E_{{\underline{a}}}{}^{a},\qquad \chi _{i}=2\,e^{-U}\ell
_{i}.  \label{FOF}
\end{equation}

Due to the Einstein equations~(\ref{Einst2}), the functions entering the
right hand sides of Eqs.~(\ref{FOF}) must satisfy the algebraic constraint
\begin{equation}
\hat{V}_{ij}=W_{i}W_{j}+\delta_{{\underline{a}}\,\bar{\underline{a}}}\Pi
_{(i}^{{\underline{a}}}\overline{\Pi }_{j)}^{\bar{\underline{a}}}+\ell
_{i}\ell _{j},  \label{alg-constr-1}
\end{equation}%
expressing the relation between $\hat{V}_{ij}$ and the $D=3$ \textit{%
Cartesian fake superpotential} $W_{i}$, and generalizing the relation~\cite%
{CD-1}%
\begin{equation}
V_{BH}=W^{2}+4G^{a\overline{a}}\partial _{a}W\overline{\partial }_{\overline{%
a}}W,
\end{equation}%
to which it reduces in the limit\footnote{%
Within the same limit, the first order constraint~(\ref{EinstConstr})
reduces to the Hamiltonian constraint given by Eq.~(11) (with $c=0$) of~\cite%
{Ferrara:1997tw}.}~(\ref{1-ctr})-(\ref{1-ctr-2}).

For later convenience, let us define the \textit{complex} $D=3$ Cartesian
vector
\begin{equation}
\mathcal{W}_{i}\equiv W_{i}+i\ell _{i},  \label{complexRot}
\end{equation}%
in terms of which the algebraic constraint~(\ref{alg-constr-1}) can be
recast as\footnote{%
For simplicity's sake, starting from below~(\ref{Einst3}) we will refrain
from underlining the scalar flat indices. In presence of curved indices, we
hope the distinction will be clear from the context.} (\textit{cfr.}~(\ref%
{Ress-1}))
\begin{equation}
\hat{V}_{ij}=\mathcal{W}_{(i}\overline{\mathcal{W}}_{j)}+\delta_{{\underline{%
a}}\,\bar{\underline{a}}}\Pi _{(i}^{{\underline{a}}}\overline{\Pi }_{j)}^{%
\bar{\underline{a}}}.  \label{Einst3}
\end{equation}%
We anticipate that the complex vector~$\mathcal{W}_{i}$ will play an
important role in integrating Maxwell Eqs.~(\ref{maxwellIntegrated}).

From now on, in order to avoid overloading the formulae, we omit underlining
the flat indices.

\subsection{\label{Constr-Flow}Construction of Flows}

We are now\ going to determine the first order flow-governing functions~$%
W_{i}$,~$\ell _{i}$ (or~$\mathcal{W}_{i}$) and~$\Pi _{i}^{a}$.

A crucial step in constructing such flow-governing functions is to use their
expression in the BPS case, and then perform a suitable \textit{flipping} of
some (linear combinations of) electromagnetic charges.

\subsubsection{\label{pSec-BPS}BPS}

As it is known~\cite{CD-1,Bossard:2011kz}, in the \textit{BPS class}~\cite%
{Denef:2000nb}, at spatial infinity (\textit{i.e.}, where one may think of a
restoration of the spherical symmetry), the first order superpotential $W$
is just a combination of the ADM mass~$M$ and the NUT charge~$N$, while the
functions~$\Pi ^{a}$ are the scalar charges\footnote{%
Here, we omit the spatial index~$i$ because in the case under consideration
the spherical symmetry reduces the number of independent components of a
Cartesian vector to one.}
\begin{equation}
W=\text{Re}(M+iN),\qquad \Pi ^{a}=\pi ^{a}.  \label{WPi}
\end{equation}%
In turn, at spatial infinity it holds
\begin{equation}
M+iN=Z,\qquad \pi ^{a}=\overline{Z}_{\overline{a}},  \label{MN}
\end{equation}%
and thus~(\ref{WPi}) can be rewritten as:
\begin{equation}
W=\text{Re}\,Z,\qquad \Pi ^{a}=\overline{Z}_{\overline{a}}.  \label{fofBPS1}
\end{equation}

In order to proceed further, we will now specialize our treatment to the $%
N=2 $, $D=4$ $stu$ model~\cite{STU}, which can also be regarded as a common
sector of $D=4$ supergravity theories with rank-$3$ symmetric scalar
manifolds.

Within this model, in order to restore manifest duality covariance in~(\ref%
{fofBPS1}), one has to consider the proper action of the compact symmetry $%
\left[ U(1)\right] ^{4}=H_{4}\times U(1)$, where $H_{4}$ is the stabilizer
of the completely factorized $D=4$ scalar manifold $\left[ SL(2,\mathbb{R}%
)/U(1)\right] ^{3}$, or equivalently the maximal compact subgroup (\textit{%
mcs}) of the $D=4$ generalized electric-magnetic ($U$-)duality group $G_{4}=%
\left[ SL(2,\mathbb{R})\right] ^{3}$, while the commuting $U(1)$ is the
\textit{mcs} of the $SL(2,\mathbb{R})$ Ehlers symmetry determined by the
reduction to tree dimensions. As discussed in~\cite{Bossard:2011kz}, under
the resulting~$[U(1)]^{4}$ symmetry, the central charge and its derivatives
transform as follows\footnote{%
The phases $\alpha _{0}$ and $\alpha _{a}$ ($a=1,2,3$) are not directly
related to the Ehlers $U(1)$ and to the three $U(1)$'s in $H_{4}$, but
rather they are a linear combinations of the corresponding phases thereof;
for further detail, see~\cite{Bossard:2011kz}.}
\begin{equation}
Z\rightarrow e^{ -\frac{i}{2}\,( \alpha _{0}-\sum\limits_{a}\alpha _{a})}
Z,\qquad Z_{a}\rightarrow e^{ -\frac{i}{2}\,( \alpha _{0}-\alpha
_{a}+\sum\limits_{\makebox[0.8em]{\tiny$b\neq a$}}\alpha _{b})} Z_{a},
\label{CCU1}
\end{equation}%
while the ADM mass~$M$, NUT charge~$N$ and scalar charges~$\pi ^{a}$
transform as
\begin{equation}
M + i N \to e^{\frac i2 (\alpha_0 + \sum\limits_a \alpha_a)} (M + i N) ,
\qquad \pi^a \to e^{-\frac i2 \left(\strut\right.\alpha_0 + \alpha_a -
\sum\limits_{\makebox[0.8em]{\tiny$b\neq a$}}\alpha_b\left.\strut\right)}%
\pi^a.
\end{equation}%
Therefore, in order to restore manifest covariance in Eq.~(\ref{fofBPS1}),
one applies transformations~(\ref{CCU1}) on the left and right hand sides of
Eq.~(\ref{MN}):
\begin{equation}
e^{\frac{i}{2}\left( \alpha _{0}+\sum\limits_{a}\alpha _{a}\right)
}(M+iN)=e^{-\frac{i}{2}\left( \alpha _{0}-\sum\limits_{a}\alpha _{a}\right)
}Z,\qquad e^{-\frac{i}{2}\strut \left( \alpha _{0}+\alpha _{a}-\sum\limits_{%
\makebox[0.8em]{\tiny$b\neq a$}}\alpha _{b}\right) }\pi ^{a}=e^{\frac{i}{2}%
\left( \alpha _{0}-\alpha _{a}+\sum\limits_{\makebox[0.8em]{\tiny$b\neq a$}%
}\alpha _{b}\right) }\bar{Z}_{\overline{a}};
\end{equation}%
by some trivial algebra, and taking into account Eq.~(\ref{WPi}), one
achieves the following result, depending only on the phase $\alpha _{0}$:
\begin{equation}
W=\text{Re}\left( e^{-i\alpha _{0}}Z\right) ,\qquad \Pi ^{a}=e^{i\alpha _{0}}%
\bar{Z}_{\overline{a}}.  \label{BPS}
\end{equation}

Although Eqs.~(\ref{BPS}) were obtained in~\cite{Bossard:2011kz} at spatial
infinity only, we will postulate that they are valid not only at the spatial
infinity, but \textit{all along the corresponding whole scalar flow} (with
generally broken spherical symmetry), and we will study the consequences of
this approach in the next Sections. Our approach then justifies the
following \textit{``Cartesian generalization''} of~(\ref{BPS}):
\begin{equation}
W_{i}=\text{Re}\left( e^{-i\alpha _{0}}\hat{Z}_{i}\right) ,\qquad \Pi
_{i}^{a}=e^{i\alpha _{0}}\hat{\overline{Z}}_{\bar{a}\,i},  \label{WPiBPS}
\end{equation}%
with~$\hat{Z}_{i}$ and~$\hat{Z}_{a\,i}$ defined in~(\ref{Zhat}).
Consequently, the function~$\ell _{i}$, and hence~$\chi _{i}$, is easily
deduced from the Einstein constraint~(\ref{Einst3}):
\begin{equation}
\ell _{i}=\text{Im}\left( e^{-i\alpha _{0}}\hat{Z}_{i}\right) \qquad
\Rightarrow \qquad \chi _{i}=2e^{-U}\text{Im}\left( e^{-i\alpha _{0}}\hat{Z}%
_{i}\right) .  \label{deffBPS}
\end{equation}%
Therefore, one realizes that the complex vector~$M+iN$ (composed by the ADM
mass~$M$ and the NUT charge~$N$) can be ``prolonged'' all along the scalar
flow by the Cartesian $D=3$ vector~$\mathcal{W}_{i}$~(\ref{complexRot}),
which in this case reads
\begin{equation}
\mathcal{W}_{i}=W_{i}+i\ell _{i}=e^{-i\alpha _{0}}\hat{Z}_{i}.
\end{equation}
The phase~$\alpha_0$ becomes a dynamical field and the consistency of the
first order equations require that it satisfy the following equation
\begin{equation*}
\partial_i \alpha_0 = e^U \text{Im} (e^{-i\alpha_0} \hat Z_i) - \text{Im}
(\partial_a K \partial_i z^a).
\end{equation*}

It is here worth stressing that Eqs.~(\ref{deffBPS}) resemble the ones
obtained in~\cite{Denef:2000nb}, but there is an important difference: they
are indeed expressed in terms of~$\hat{Z}_{i}$, instead of~$Z_{i}$. The very
definition~(\ref{Zhat}) of the $D=3$ \textit{Cartesian }$N=2$\textit{\
central charge}~$\hat{Z}_{i}$ involves terms proportional to the vector~$%
\chi _{i}$, thus Eq.~(\ref{deffBPS}) can actually be solved for~$\chi _{i}$
itself, achieving the following expression:
\begin{equation}
\chi _{i}=\frac{2e^{-U}\text{Im}\left[ e^{-i\alpha _{0}}Z_{i}\right] }{%
1-2e^{-U}\text{Im}\left[ e^{-i\alpha _{0}}\langle b,V\rangle \right] }.
\end{equation}%
Then, the integration of Maxwell Eqs.~(\ref{maxwellIntegrated}) exactly
implies the result of~\cite{Denef:2000nb}. In order to achieve this, one is
hinted by the fact that expression~(\ref{WPiBPS}) can be recast in the
following form:
\begin{equation*}
W_{i}=\text{Re}\,\langle \hat{H}_{i},e^{-i\alpha _{0}}V\rangle ,
\end{equation*}%
and one can thus check that
\begin{equation}
b^{\alpha }=2e^{U}\text{Re}\left( e^{-i\alpha _{0}}V^{\alpha }\right)
\label{emBPS}
\end{equation}%
satisfies the Maxwell equations~(\ref{maxwellIntegrated}), thus yielding the
following flow-defining functions:
\begin{equation}
W_{i}=\text{Re}\left( e^{-i\alpha _{0}}Z_{i}\right) ,\qquad \Pi
_{i}^{a}=e^{i\alpha _{0}}\bar{Z}_{\bar{a}\,i},\qquad \chi _{i}=-2e^{-U}\text{%
Im}\left( e^{-i\alpha _{0}}Z_{i}\right) ,  \label{FO-BPS-class}
\end{equation}%
which govern the first-order formulation of scalar flows of the \textit{BPS
class} of multi-centered BHs~\cite{Denef:2000nb}.

\subsubsection{\label{pSec-almost-BPS}Almost BPS}

In order to construct a first order formalism for non-BPS flows, we will
exploit suitable \textit{charge flippings} at spatial infinity, and then
consistently extend them all along the flow. The relevant flippings of
charges have been derived in~\cite{Bossard:2011kz} by exploiting an analysis
of the relevant nilpotent orbits of $SO(4,4)$, which is the $D=3$ duality
group of the $stu$ model.

We start and consider the central charge~$Z$ and its flat derivatives~$Z_{a}$
at spatial infinity. Without loss of generality (by $H_{4}=U(1)^{3}$
duality), the asymptotical values of the scalar fields~$z^{a}$ can be set to
\begin{equation}
z^{a}=-i.
\end{equation}%
Therefore, at spatial infinity the central charge along with its flat K\"{a}%
hler-covariant derivatives respectively read%
\begin{equation}
Z=\frac{1}{2\sqrt{2}}\makebox[0.5em]{$\left[\rule{0pt}{1.1em}\right.$}%
q_{0}+ip^{0}-\sum_{a}\,(p^{a}-iq_{a})\makebox[0.5em]{$\left.%
\rule{0pt}{1.1em}\right]$},\qquad Z_{a}=\frac{1}{2\sqrt{2}}%
\makebox[0.5em]{$\left[\rule{0pt}{1.1em}\right.$}q_{0}-ip^{0}+\sum\limits_{%
\makebox[0.8em]{\tiny$b\neq a$}}(p^{b}+iq_{b})-p^{a}+iq_{a}%
\makebox[0.5em]{$\left.\rule{0pt}{1.1em}\right]$}.  \label{ZZ-1}
\end{equation}%
Instead of dealing with the charges~$(p^{0},p^{a},q_{0},q_{a})$, let us
define a basis $(D^{0},D^{a},D_{0},D_{a})$ whose interpretation in terms of $%
D$-brane charges is given below~\cite{Bossard:2011kz}:
\begin{equation}
\begin{array}{lll}
\displaystyle\text{D6}:D^{0}\equiv \frac{1}{2}\makebox[0.5em]{$\left[%
\rule{0pt}{1.1em}\right.$}p^{0}+\sum_{a}q_{a}\makebox[0.5em]{$\left.%
\rule{0pt}{1.1em}\right]$}, & \displaystyle\quad  & \displaystyle\text{D4}%
:D^{a}\equiv p^{a}, \\
\displaystyle\text{D2}:D_{a}\equiv \frac{1}{2}\makebox[0.5em]{$\left[%
\rule{0pt}{1.1em}\right.$}p^{0}+q_{a}-\sum_{b\neq a}q_{b}\makebox[0.5em]{$%
\left.\rule{0pt}{1.1em}\right]$}, & \displaystyle\quad  & \displaystyle\text{%
D0}:D_{0}\equiv q_{0};%
\end{array}
\label{D-charges}
\end{equation}%
this is a more convenient basis on which one can act with the charge
flipping. In terms of $D$-charges~(\ref{D-charges}),~(\ref{ZZ-1}) can be
rewritten as
\begin{equation}
\begin{array}{l}
\displaystyle Z=\frac{1}{2\sqrt{2}}\makebox[0.5em]{$\left[\rule{0pt}{1.1em}%
\right.$}D_{0}-iD^{0}+\sum_{a}\,(D^{a}+iD_{a})\makebox[0.5em]{$\left.%
\rule{0pt}{1.1em}\right]$}, \\
\displaystyle Z_{a}=\frac{1}{2\sqrt{2}}\makebox[0.5em]{$\left[%
\rule{0pt}{1.1em}\right.$}D_{0}-iD^{0}-\sum_{b\neq
a}\,(D^{b}+iD_{b})+D^{a}+iD_{a}\makebox[0.5em]{$\left.\rule{0pt}{1.1em}%
\right]$},%
\end{array}
\label{Z2D}
\end{equation}%
or equivalently, inverting in terms of the $D$-charges:
\begin{equation}
\begin{array}{l}
\displaystyle D_{0}=\frac{1}{\sqrt{2}}\,\text{Re}\makebox[0.5em]{$\left[%
\rule{0pt}{1.1em}\right.$}Z+\sum_{a}Z_{a}\makebox[0.5em]{$\left.%
\rule{0pt}{1.1em}\right]$},\quad \displaystyle D_{a}=\frac{1}{\sqrt{2}}\,%
\text{Im}\makebox[0.5em]{$\left[\rule{0pt}{1.1em}\right.$}%
Z+Z_{a}-\sum_{b\neq a}Z_{b}\makebox[0.5em]{$\left.\rule{0pt}{1.1em}\right]$},
\\
\displaystyle D^{0}=-\frac{1}{\sqrt{2}}\,\text{Im}\makebox[0.5em]{$\left[%
\rule{0pt}{1.1em}\right.$}Z+\sum_{a}Z_{a}\makebox[0.5em]{$\left.%
\rule{0pt}{1.1em}\right]$},\quad \displaystyle D^{a}=\frac{1}{\sqrt{2}}\,%
\text{Re}\makebox[0.5em]{$\left[\rule{0pt}{1.1em}\right.$}%
Z+Z_{a}-\sum_{b\neq a}Z_{b}\makebox[0.5em]{$\left.\rule{0pt}{1.1em}\right]$}.%
\end{array}
\label{D2Z}
\end{equation}%
\medskip

Now, consistent with the analysis of~\cite{Bossard:2011kz}, let us flip the
sign of the brane charge $D_{0}$~(\ref{D-charges}) in~(\ref{Z2D}), obtaining%
\begin{equation}
\begin{array}{l}
\displaystyle\widetilde{Z}\equiv \left. Z\right\vert _{D_{0}\rightarrow
-D_{0}}=\frac{1}{2\sqrt{2}}\makebox[0.5em]{$\left[\rule{0pt}{1.1em}\right.$}
-D_{0}-iD^{0}+\sum_{a}\,(D^{a}+iD_{a})\makebox[0.5em]{$\left.%
\rule{0pt}{1.1em}\right]$} , \\
\displaystyle \widetilde{Z}_{a}\equiv \left. Z_{a}\right\vert
_{D_{0}\rightarrow -D_{0}}=\frac{1}{2\sqrt{2}}\makebox[0.5em]{$\left[%
\rule{0pt}{1.1em}\right.$} -D_{0}-iD^{0}-\sum_{b\neq
a}\,(D^{b}+iD_{b})+D^{a}+iD_{a}\makebox[0.5em]{$\left.\rule{0pt}{1.1em}%
\right]$} ;%
\end{array}
\label{Z2D2}
\end{equation}%
by plugging the expressions~(\ref{D2Z}) into~(\ref{Z2D2}), one achieves the
following result:
\begin{equation}
\widetilde{Z}=\frac{1}{4}\,\makebox[0.5em]{$\left[\rule{0pt}{1.1em}\right.$}
3Z-\bar{Z}-2\,\text{Re} \sum_{a}Z_{a} \makebox[0.5em]{$\left.%
\rule{0pt}{1.1em}\right]$} ,\qquad \widetilde{Z}_{a}=-\frac{1}{2}%
\makebox[0.5em]{$\left[\rule{0pt}{1.1em}\right.$} \text{Re}Z+\text{Re}%
\sum_{b}Z_{b}-2Z_{a}\makebox[0.5em]{$\left.\rule{0pt}{1.1em}\right]$},
\end{equation}%
which can then be plugged into the right hand side of Eqs.~(\ref{MN}). By
restoring manifest $\left( U(1)\right) ^{4}$ covariance in the way described
above, and extending these relations along the whole flow and not only at
spatial infinity, one obtains the following expressions, explicitly
depending on \textit{all four} phases $\alpha _{0}$ and $\alpha _{a}$'s:
\begin{eqnarray}
&&%
\begin{array}{l}
W_{i}=\frac{1}{4}\text{Re}\rule{0pt}{1em}\left[ e^{-i\alpha _{0}}\left(
3-e^{i(\alpha _{0}+\sum\limits_{a}\alpha _{a})}\right) \hat{Z}_{i}-\left(
1+e^{-i(\alpha _{0}+\sum\limits_{a}\alpha _{a})}\right)
\sum\limits_{a}e^{i\alpha _{a}}\hat{Z}_{a\,i}\right] \rule{0pt}{1em}; \\
\\
\Pi _{i}^{a}=\frac{1}{2}e^{i\alpha _{0}}\hat{\overline{Z}}_{a\,i}-\frac{1}{2}%
e^{\frac{i}{2}(\alpha _{0}+\alpha _{a}-\sum\limits_{b\neq a}\alpha _{b})}%
\text{Re}\rule{0pt}{1em}\left( e^{-\frac{i}{2}(\alpha
_{0}-\sum\limits_{b}\alpha _{b})}\hat{Z}_{i}\right) \rule{0pt}{1em} \\
~~~~-\frac{i}{2}e^{\frac{i}{2}(\alpha _{0}+\alpha _{a}-\sum\limits_{b\neq
a}\alpha _{b})}\text{Im}\rule{0pt}{1em}\left[ \sum\limits_{b}e^{-\frac{i}{2}%
(\alpha _{0}-\alpha _{b}+\sum\limits_{c\neq b}\alpha _{c})}\hat{Z}_{b\,i}%
\right] \rule{0pt}{1em}.%
\end{array}
\notag \\
&&  \label{defAlmost}
\end{eqnarray}%
This case corresponds to the so-called \textit{almost BPS} class~\cite%
{Goldstein-Katmadas,Bossard:2011kz}, whose first order formulation will be
studied in Sec. \ref{sectionAlmost}.

\subsubsection{\label{pSec-composite-nBPS}Composite Non-BPS}

Another possibility is to flip not only $D_{0}$, but also $D_{a}$ brane
charges~(\ref{D-charges}) as well, namely:
\begin{equation}
D_{0}\rightarrow -D_{0},\qquad D_{a}\rightarrow -D_{a}.  \label{sign-flip-2}
\end{equation}%
By applying the very same procedure described above for the sign flip~(\ref%
{sign-flip-2}) , one finally gets the following expressions, explicitly
depending on only\textit{\ three} phases $\alpha _{a}$'s:
\begin{equation}
\begin{array}{l}
\displaystyle W_{i}=\frac{1}{2}\text{Re}\rule{0pt}{1em}\left(
e^{i\sum\limits_{a}\alpha _{a}}\hat{Z}_{i}-\sum\limits_{a}e^{i\alpha _{a}}%
\hat{Z}_{a\,i}\right) \rule{0pt}{1em}; \\[0.5em]
\displaystyle\Pi _{i}^{a}=-\frac{1}{2}\rule{0pt}{1em}\left[ e^{i\alpha _{a}}%
\hat{Z}_{i}-e^{i(\alpha _{a}-\sum\limits_{b\neq a}\alpha _{b})}\hat{Z}%
_{a\,i}+\sum\limits_{\makebox[0.8em]{\tiny$b\neq a$}}e^{-i\alpha _{b}}\hat{Z}%
_{b\,i}\right] \rule{0pt}{1em}.%
\end{array}
\label{fofComp}
\end{equation}

This case corresponds to the so-called \textit{composite non-BPS} class~\cite%
{Bossard:2011kz}, whose first order formulation will be studied in Sec. \ref%
{sectionComposite}.

\section{\label{sectionComposite}Composite Non-BPS Class}

We will start analyzing the \textit{composite non-BPS}~\cite{Bossard:2011kz}
class~(\ref{fofComp}), its structure being relatively simpler than the one
of the almost-BPS class.

In the previous Section, we constructed the functions~$W_{i}$ and~$\Pi
_{i}^{a}$~(\ref{fofComp}), governing the first order flow Eqs.~(\ref{FO-1}).
From Einstein Eqs.~(\ref{Einst2}), one can deduce the following expression
for the remaining flow-governing function, namely the $D=3$ Cartesian
rotation vector $\ell _{i}$:
\begin{equation}
\ell _{i}=-\frac{1}{2}\text{Im}\left[ e^{i\sum\limits_{a}\alpha _{a}}\hat{Z}%
_{i}-\sum_{a}e^{i\alpha _{a}}\hat{Z}_{a\,i}\right] \rule{0pt}{1em}.
\label{deffalmost}
\end{equation}

As it can be realized by looking at previous equations, the various $\alpha $%
-phases play an important role in the first order formalism for the
multi-centered scalar flows. Nevertheless, their own dynamics has not been
concerned so far. In the case This can be uniquely determined if one imposes
that after differentiation of the first order equations~(\ref{FOF}),~(\ref%
{fofComp}) and~(\ref{deffalmost}), the second order ones~(\ref{eom3}) are
obtained. This requirement yields the following three partial differential
Eqs. for the three phases $\alpha _{a}$'s ($a=1,2,3$) in the composite
non-BPS multi-centered class\footnote{%
It is worth noticing that both sets~(\ref{derPhase}) and~(\ref%
{solDPhasesAlmost}) of partial differential Eqs. for the phases
(respectively in the composite non-BPS e almost BPS classes) exhibit a
manifest $stu$ \textit{triality} symmetry~\cite{STU}.}:
\begin{equation}
\begin{array}{l}
\displaystyle\partial _{i}\alpha _{a}=\frac{1}{2}\,e^{U}\,\text{Im}\left[
\rule{0pt}{1em}\makebox[0.5em]{$\left(\rule{0pt}{1em}\right.$}
e^{i\sum\limits_{b}\alpha _{b}}+e^{i\alpha _{a}}-\sum\limits_{b\neq
a}e^{i\alpha _{b}}\makebox[0.5em]{$\left.\rule{0pt}{1em}\right)$} \rule%
{0pt}{1em}\hat{Z}_{i}+2e^{i\alpha _{a}}\hat{Z}_{a\,i}-2e^{-i\alpha
_{a}}\sum\limits_{b\neq a}\hat{Z}_{b\,i}\right. \\
\displaystyle\phantom{\partial _{i}\alpha _{a}=\frac12}\left. +\,e^{-i\alpha
_{a}}\rule{0pt}{1em}\makebox[0.5em]{$\left(\rule{0pt}{1em}\right.$}
e^{-i\sum\limits_{b}\alpha _{b}}-e^{i\alpha _{a}}+\sum\limits_{b\neq
a}e^{-i\alpha _{b}}\makebox[0.5em]{$\left.\rule{0pt}{1em}\right)$} %
\makebox[0.5em]{$\left(\rule{0pt}{1em}\right.$} -e^{i\alpha _{a}}\hat{Z}%
_{a\,i}+\sum\limits_{b\neq a}e^{i\alpha _{b}}\hat{Z}_{b\,i}%
\makebox[0.5em]{$\left.\rule{0pt}{1em}\right)$}\right].%
\end{array}
\label{derPhase}
\end{equation}%
Thus, one can conclude that Eqs.~(\ref{FOF}), (\ref{fofComp}), (\ref%
{deffalmost}), and~(\ref{derPhase}) -- together with Maxwell Eqs.~(\ref%
{maxwellIntegrated}) -- constitute the \textit{first order formalism for
composite non-BPS multi-centered BH solutions} in $N=2$, $D=4$ $stu$ model.

Let us now integrate further Maxwell Eqs.~(\ref{maxwellIntegrated}) by
exploiting the hint described in Sec. \ref{pSec-BPS} for the BPS flow.
Namely, one can easily notice that for the composite non-BPS class the
complex $D=3$ Cartesian vector~$\mathcal{W}_{i}$~(\ref{complexRot}) can be
rewritten as
\begin{equation}
\mathcal{W}_{i}=\langle \hat{H}_{i},T\rangle ,  \label{WfasReIm}
\end{equation}%
where the complex symplectic vector
\begin{equation}
T^{\alpha }\equiv \frac{1}{2}\rule{0pt}{1em}\left(
e^{-i\sum\limits_{a}\alpha _{a}}\bar{V}^{\alpha }-\sum_{a}e^{-i\alpha _{a}}%
\bar{V}_{\bar{a}}^{\alpha }\right) \rule{0pt}{1em}  \label{Talmost}
\end{equation}%
was defined. After a little algebra, one also finds that electromagnetic
potential~$b$~(\ref{b}) enjoys an expression similar to the one holding for
the BPS class~(\ref{emBPS}), namely:
\begin{equation}
b^{\alpha }=2\,e^{U}\text{Re}\rule{0pt}{1em}\left[ T^{\alpha }\left(
1-iBe^{2U}\right) \right] \rule{0pt}{1em},  \label{balmost}
\end{equation}%
for some function~$B$, which can be considered as the $D=5$\textit{\
rotational contribution} to the electromagnetic potential.

As it occurred in the BPS multi-centered class treated in Sec. \ref{pSec-BPS}%
, also in the composite non-BPS class Eq.~(\ref{deffalmost}) contains~$\chi
_{i}$ in both sides. Remarkably, in this class the dependence on $\chi _{i}$
drops out when replacing~(\ref{balmost}) into~(\ref{deffalmost}); as a
consequence, Eq.~(\ref{deffalmost}) turns into the following algebraic
constraint:
\begin{equation}
\text{Im}\left( e^{i\sum\limits_{a}\alpha _{a}}Z_{i}-\sum_{a}e^{i\alpha
_{a}}Z_{a\,i}\right) =0,  \label{constrComp}
\end{equation}%
and the $D=3$ Cartesian rotation vector $\chi _{i}$ can be computed to read
\begin{equation}
\chi _{i}=\partial _{i}B+Be^{U}\,\text{Re}\left( \strut
3\,e^{i\sum\limits_{a}\alpha _{a}}Z_{i}-\sum_{a}e^{i\alpha
_{a}}Z_{a\,i}\right) -2e^{-U}\text{Im}\,\left( e^{i\sum\limits_{a}\alpha
_{a}}Z_{i}\right) ,  \label{deffcomposit2}
\end{equation}%
where the function~$B$ occurred for the first time in~(\ref{balmost}), and
it should be such that $\chi _{i}$ is divergenceless (\textit{cfr.}~(\ref%
{divergenless})).

The counting of the two ``flat'' directions along the (non-BPS, ``large'')
flows in the $stu$ model~\cite{Bellucci:2008sv,FM-2} is retrieved by
considering that in this class the tree partial differential Eqs.~(\ref%
{derPhase}) for the tree phases $\alpha _{a}$'s are supplemented by the
algebraic constraint~(\ref{constrComp}) (see also the treatment of next
Subsection).

\subsection{Single-Centered Solutions}

Consistent with~\cite{Bossard:2011kz} (see also \textit{e.g.} the
two-centered analysis in~\cite{CFMY-Small-1}), we will now show that the
well known case of a single-centered non-rotating non-BPS BH can be obtained
by performing a suitable limit in the multi-centered class under
consideration.

We consider the single-centered limit~(\ref{1-ctr})-(\ref{harmonicH}), in
which the divergenceless vector~$H_{i}^{\alpha }$ introduced in~(\ref%
{maxwellIntegrated}) is realized as a derivative of an harmonic function~$%
H^{\alpha }$ defined in~(\ref{harmonicH}):
\begin{equation}
H_{i}^{\alpha }=\partial _{i}H^{\alpha },\qquad \partial _{i}\partial
_{i}H^{\alpha }=0.
\end{equation}%
Before proceeding further, it is worth clarifying here the meaning of the
constraint~(\ref{constrComp}). With such a purpose, let us introduce the
three real values
\begin{equation}
\lambda ^{a}\equiv \frac{\text{Im}\,(e^{i\beta _{a}}\,z^{a})}{\text{Im}%
\,\left( e^{i\beta _{a}}\right) },\qquad \beta _{a}\equiv \frac{1}{2}%
\sum\limits_{b\neq a}\alpha _{b}  \label{intMot}
\end{equation}%
and construct an~$stu$-like symplectic section%
\begin{equation}
V_{\lambda }^{\alpha }\equiv (1,\lambda ^{1},\lambda ^{2},\lambda
^{3},-\lambda ^{1}\lambda ^{2}\lambda ^{3},\lambda ^{2}\lambda ^{3},\lambda
^{1}\lambda ^{3},\lambda ^{1}\lambda ^{2})^{T}.
\end{equation}%
This allows the following rather elegant rewriting of the constraint~(\ref%
{constrComp}):
\begin{equation}
\langle P,V_{\lambda }\rangle =0.  \label{intMotConstr}
\end{equation}%
Note that the three~$\lambda ^{a}$'s satisfy the single constraint~(\ref%
{intMotConstr}), and therefore only two of them are independent; this
reflects the existence of two ``flat'' directions along the non-rotating
single-centered $stu$ non-BPS flow~\cite{Bellucci:2008sv,FM-2}. A noteworthy
feature of $\lambda ^{a}$'s is that they are in fact \textit{integrals of
motion }in $D=3$ flat space; indeed, from the above Eqs. one can check that
\begin{equation}
\partial _{i}\lambda ^{a}=0.
\end{equation}%
By inverting the definitions~(\ref{intMot}), one can thus express the three
phases~$\alpha _{a}$'s in terms of the scalar fields~$z^{a}$ and of the
integrals of motion $\lambda ^{a}$'s themselves:
\begin{equation}
e^{i\sum\limits_{b\neq a}\alpha _{b}}=\frac{\lambda ^{a}-\overline{z}^{\bar{a%
}}}{\lambda ^{a}-z^{a}}.  \label{solPhasesComp}
\end{equation}%
In the case of spherical symmetry, each of the flow-defining $D=3$ Cartesian
vectors~$W_{i}$ and~$\Pi _{i}^{a}$ can actually be reduced to only one
independent function:
\begin{equation}
W_{i}=\partial _{i}\tau \,W,\qquad \Pi _{i}^{a}=\partial _{i}\tau \,\Pi ^{a},
\end{equation}%
where $\tau $ has been defined in~(\ref{harmonicH}); therefore, the spatial
vector index $i$ can formally be neglected.

By plugging the phases~(\ref{solPhasesComp}) into the Eqs.~(\ref{fofComp}),
one retrieves the known expression for the non-BPS fake superpotential~$%
W_{nBPS}$ for a non-rotating single-centered non-BPS BH in~$stu$ model (with
non-vanishing $Z$ at the horizon)~\cite{Bellucci:2008sv}; in order to
achieve a complete agreement in notations, one should parametrize the
integrals of motion as
\begin{equation}
\lambda ^{a}\equiv \frac{e^{\mathbf{\alpha }_{a}}\nu \xi _{a}+\rho _{a}}{e^{%
\mathbf{\alpha }_{a}}\nu -1},
\end{equation}%
where the~$\mathbf{\alpha }_{a}$'s are some constant (not to be confused
with the phases $\alpha _{a}$'s) satisfying~$\sum\limits_{a}\mathbf{\alpha }%
_{a}=0$.

It should be here remarked that in the non-rotating ($\chi _{i}=0$)
single-centered non-BPS BH solution supported by the Kaluza-Klein $\left(
p^{0},q_{0}\right) $-configuration~\cite{GLS-1,Bellucci:2008sv}
\begin{eqnarray}
e^{-4U} &=&H_{0}H_{1}H_{2}H_{3}-B^{2},\quad z^{a}=\frac{A_{a}^{-}-2ie^{-2U}}{%
A_{a}^{+}-2B}; \\
B &=&\mbox{const},\qquad A_{a}^{\pm }=\frac{H_{1}H_{2}H_{3}}{H_{a}}\pm
H_{0}H_{a},
\end{eqnarray}%
the function~$\chi _{i}$, defined by eq.~(\ref{deffcomposit2}), vanishes. On
the other hand, if the function~$B$ is not a constant, then it turns out to
read
\begin{equation*}
B=b+J\frac{\cos \theta }{r^{2}},
\end{equation*}%
where~$J$ is an angular momentum; this corresponds to the Rasheed-Larsen
rotating ($\chi _{i}\neq 0$) single-centered non-BPS BH~\cite%
{Rasheed:1995zv, Larsen:1999pp}.

\subsection{Multi-Centered Solutions}

\subsubsection{Electric Configuration}

In~\cite{Bossard:2011kz}, it was found that for the composite non-BPS class
there exists a particular solution
\begin{equation}
\displaystyle e^{-4U}=4\,V_{0}H_{1}H_{2}H_{3}-B^{2},\quad z^{a}=\frac{1}{2}\,%
\frac{2K_{a}\frac{H_{1}H_{2}H_{3}}{H_{a}}-B-i\,e^{-2U}}{K_{a}^{2}\frac{%
H_{1}H_{2}H_{3}}{H_{a}}+H_{a}V_{0}-BK_{a}},  \label{pre-1}
\end{equation}%
defined in terms of functions~$V_{0},H_{a}$ and~$K_{a}$ satisfying the
following equations:
\begin{equation}
\begin{array}{l}
\displaystyle\partial _{i}V_{0}=H_{i}^{0}-K_{a}H_{i}^{a}+\frac{1}{2}%
\left\vert \epsilon ^{abc}\right\vert K_{a}K_{b}H_{c\,i}, \\[0.5em]
\displaystyle\partial _{i}K_{a}=\frac{H_{a}}{2H_{1}H_{2}H_{3}}%
\makebox[0.5em]{$\left[\rule{0pt}{1.1em}\right.$}\sum\limits_{%
\makebox[0.8em]{\tiny$b\neq a$}}H_{a}H_{i}^{a}-H_{b}H_{i}^{b}+\left\vert
\epsilon ^{bcd}\right\vert \,K_{b}H_{c}H_{di}-2H_{a}\sum\limits_{{\tiny
\makebox[0.5em]{$\ba{c}b\neq a\neq
c\\b\neq c\ea$}}}K_{b}H_{c\,i}\makebox[0.5em]{$\left.\rule{0pt}{1.1em}%
\right]$}, \\
\displaystyle\chi _{i}=\partial _{i}B-\left\vert \epsilon ^{abc}\right\vert
\partial _{i}K_{a}H_{b}H_{c},%
\end{array}
\label{defVKLcomp}
\end{equation}%
%
%
%
%
%
%
%
%
%
%
%
%
%
%
%
%
%
%
%
%
%
in which~$H_{i}^{0}$, $H_{i}^{a}$ and $H_{ai}=\partial _{i}H_{a}$ are
components of the divergenceless symplectic vector~$H_{i}^{\alpha }$
introduced in~(\ref{maxwellIntegrated}). Without entering in the details
\textit{e.g.} of the two-centered solution, it should be remarked here that
in the composite non-BPS class the fact that $H_{i}^{0}$ and $H_{i}^{a}$ are
not harmonic implies the centers to exhibit \textit{mutually non-local}
electric-magnetic fluxes. The same holds for the magnetic solution
considered in Sec. \ref{MS}.

One can check that this solution satisfies the composite non-BPS first order
equations derived above if the phases~$\alpha _{a}$'s are fixed as follows:
\begin{equation}
\frac{1}{2}\sum_{b\neq a}\alpha _{b}=\pi -\text{arg}\, z^{a}.
\label{phasesBossard}
\end{equation}%
The condition~(\ref{phasesBossard}) is a particular solution to the partial
differential equations~(\ref{derPhase}), and it actually fixes the two
non-BPS ``flat'' directions, whose presence cannot thus be recognized in the
explicit form~(\ref{pre-1})-(\ref{defVKLcomp}) of this solution. Moreover,
it should be noted that~(\ref{phasesBossard}) yields the vanishing of the
remaining components (not entering~(\ref{defVKLcomp})) of the divergenceless
symplectic vector~$H_{i}^{\alpha }$:
\begin{equation}
H_{0\,i}=0.  \label{z-1}
\end{equation}%
Physically, this means that the graviphoton electric charge~$q_{0}$
pertaining to \textit{each} BH center should vanish. However, in the
two-centered case, it can be checked that in general the charge vectors
pertaining to each of the two centers are \textit{mutually non-local}, even
if there are no solutions in this class in which all two-centered duality
invariants~\cite{FMOSY-1} are independent (also \textit{cfr.} the analysis
in~\cite{CFMY-Small-1}).

\paragraph{Multi-Centered Integrals of Motion}

Although introduced for single-centered BH solutions, the definition~(\ref%
{intMot}) of integrals of the $D=3$ motion still holds for multi-centered
BHs. In this case, the constraint~(\ref{constrComp}) can be rewritten as
\begin{equation}
\makebox[0.5em]{$\left[\rule{0pt}{1em}\right.$}\prod_{a}\,\text{Im}\,
e^{i\beta _{a}} \makebox[0.5em]{$\left.\rule{0pt}{1em}\right]$} \langle
H_{i},V_{\lambda }\rangle =0.  \label{constrCompGamma}
\end{equation}%
Assuming that the phases~$\alpha_a$ are such that the product of the
imaginary parts (recall definition~(\ref{intMot})) does not vanish, one
obtains that that the skew-symmetric symplectic product~$\langle
H_{i},V_{\lambda }\rangle $ necessarily vanishes:
\begin{equation}
\langle H_{i},V_{\lambda }\rangle =H_{0\,i}+H_{a\,i}\lambda
^{a}-3H_{i}^{(1}\lambda ^{2}\lambda ^{3)}+H_{i}^{0}\lambda ^{1}\lambda
^{2}\lambda ^{3}=0.  \label{constrIntMot2}
\end{equation}%
From the definition of integrals of the $D=3$ motion, and especially from
its inverse~(\ref{solPhasesComp}), one can see that~(\ref{phasesBossard})
holds \textit{iff} all integrals of motion equal to zero:~$\lambda ^{a}=0$, $%
\forall a=1,2,3$. In order to fulfill the constraint~(\ref{constrIntMot2}),
one has to require~$H_{0\,i}=0$, matching~(\ref{z-1}).

The product of the imaginary parts in~(\ref{constrCompGamma}) might vanish.
This happens for example in the case of the so-called magnetic
configuration, which analysis is performed below.

\subsubsection{\label{MS}Magnetic Configuration}

If all phases~$\alpha _{a}$ vanish, then the constraint~(\ref{derPhase})
turns into an identity. Despite the fact that the definition~(\ref{intMot})
of integrals of motion becomes singular, a careful analysis yields that the
constraint~(\ref{constrCompGamma}) is satisfied when
\begin{equation}
H_{i}^{0}=0.
\end{equation}%
Physically, this condition means that the graviphoton magnetic charge~$p^{0}$
of \textit{each} BH center should vanish. This case was considered in~\cite%
{Bossard-Octonionic}; therein, an explicit solution in the magnetic
configuration
\begin{equation}
e^{-4U}=4V_{0}H^{1}H^{2}H^{3}-B^{2},\quad z^{a}=K^{a}+\frac{H^{a}}{%
2H^{1}H^{2}H^{3}}\,({B-ie^{-2U}})
\end{equation}%
was found, in which the function~$V_{0}$ as well as the harmonic functions~$%
K^{a}$ satisfy the first order equations (no summation over index~$a$):
\begin{equation}
\begin{array}{l}
\displaystyle\partial _{i}V_{0}=-H_{0i}-K^{b}H_{bi}+\frac{1}{2}|\epsilon
_{bcd}|K^{b}K^{c}\partial _{i}H^{d}, \\[0.5em]
\displaystyle\partial _{i}K^{a}=\frac{H^{a}}{2H^{1}H^{2}H^{3}}%
\makebox[0.5em]{$\left[\rule{0pt}{1.1em}\right.$}H^{a}H_{ai}-\sum\limits_{b%
\neq a}H^{b}H_{bi}+|\epsilon _{bcd}|H^{b}K^{c}\partial
_{i}H^{d}-2H^{a}\sum\limits_{{\tiny
\makebox[0.5em]{$\ba{c}b\neq a\neq
c\\b\neq c\ea$}}}K^{b}H_{\,i}^{c}\makebox[0.5em]{$\left.\rule{0pt}{1.1em}%
\right]$}.%
\end{array}%
\end{equation}%
Thus, the $D=3$ Cartesian rotation vector reads
\begin{equation}
\chi _{i}=\partial _{i}B+|\epsilon _{abc}|\partial _{i}K^{a}H^{b}H^{c}.
\end{equation}%
This solution, which is a \textit{particular} solution of the composite
non-BPS first order equations, is written in terms of the functions~$H_{0i}$%
,~$H_{ai}$ and $H_{i}^{a}=\partial _{i}H^{a}$, which are components of the
divergenceless symplectic vector~$H_{i}^{\alpha }$~introduced in (\ref%
{maxwellIntegrated}).

\section{\label{sectionAlmost}Almost BPS Class}

For the \textit{almost BPS}~\cite{Goldstein-Katmadas,Bossard:2011kz} class
of flows, the first order equations are given by the general Eqs.~(\ref{FOF}%
) supplemented with the flow-defining functions~$W_{i}$ and~$\Pi _{i}^{a}$
given by Eqs.~(\ref{defAlmost}). Then, one has to satisfy the Einstein
constraint~(\ref{Einst3}), yielding the following expression for the $D=3$
Cartesian rotation vector $\ell _{i}$:
\begin{equation}
\ell _{i}=\frac{1}{4}\,\text{Im}\left[ e^{-i\alpha _{0}}\left( 3+e^{i(\alpha
_{0}+\sum\limits_{a}\alpha _{a})}\right) \hat{Z}_{i}+\left( 1-e^{-i(\alpha
_{0}+\sum\limits_{a}\alpha _{a})}\right) \sum\limits_{a}e^{i\alpha _{a}}\hat{%
Z}_{a\,i}\right] .  \label{fAlmost}
\end{equation}

As described in the Sec. \ref{sectionComposite}, if one requires the
equivalence of the first order equations to their second order ancestors,
the following four partial differential Eqs. for the four phases $\alpha
_{0} $ and $\alpha _{a}$'s are obtained:
\begin{equation}
\begin{array}{l}
\partial _{i}\alpha _{0}=e^{U}\text{Im}\,\left[ e^{-i\alpha _{0}}(\hat{Z}%
_{i}+\sum\limits_{a}\hat{Z}_{a\,i})\right] \\
~~~-\frac{1}{4}e^{U}\left[ (e^{i\sum\limits_{a}\alpha
_{a}}-\sum\limits_{a}e^{i\alpha _{a}}+e^{-i\alpha
_{0}}(\sum\limits_{a}e^{i\sum\limits_{b\neq a}\alpha _{b}}-1))\hat{Z}%
_{i}\right. \\
~~~~\left. +\sum\limits_{a}\left[ (e^{i\alpha _{a}}+e^{-i\alpha
_{0}})(1-e^{-i\sum\limits_{b\neq a}\alpha _{b}})-\sum\limits_{b\neq
a}e^{-i\alpha _{b}}(1-e^{i\alpha _{a}-i\alpha _{0}})\right] \hat{Z}_{a\,i}%
\right] , \\
\\
\partial _{i}\alpha _{a}=-\partial _{i}\alpha _{0}-\frac{1}{2}e^{U}\cos {%
\frac{\left( \alpha _{0}+\alpha _{a}\right) }{2}}\,\text{Im}\rule{0pt}{1.1em}%
\left[ e^{-\frac{i}{2}(\alpha _{0}-\alpha _{a})}\rule{0pt}{1.1em}\left[
(e^{i\sum\limits_{b\neq a}\alpha _{b}}-1)\hat{Z}_{i}\right. \right. \\
~~~\left. \left. -(e^{-i\sum\limits_{b\neq a}\alpha _{b}}+3)\hat{Z}_{a\,i}%
\right] \rule{0pt}{1.1em}+e^{-\frac{i}{2}(\alpha _{0}+\alpha
_{a})}\sum\limits_{b\neq a}(e^{i\alpha _{b}}-e^{-i\alpha _{c}})\hat{Z}_{b\,i}%
\right] \rule{0pt}{1.1em},%
\end{array}
\label{solDPhasesAlmost}
\end{equation}%
in which the index~$c$ must be different from both~$a$ and~$b$. In turn,
Eqs.~(\ref{solDPhasesAlmost}) yield a useful relation:%
\begin{equation}
\partial _{i}\makebox[0.5em]{$\left[\rule{0pt}{1.1em}\right.$} \alpha
_{0}+\sum\limits_{a}\alpha _{a}\makebox[0.5em]{$\left.\rule{0pt}{1.1em}%
\right]$} =2\,e^{U}\,\text{Re}\makebox[0.5em]{$\left[\rule{0pt}{1.1em}%
\right.$} e^{\frac{i}{2}(\alpha _{0}+\sum\limits_{a}\alpha _{a})}%
\makebox[0.5em]{$\left.\rule{0pt}{1.1em}\right]$}\, \text{Im}\left[ e^{\frac{%
i}{2}(\alpha _{0}+\sum\limits_{a}\alpha _{a})}\makebox[0.5em]{$\left(%
\rule{0pt}{1.1em}\right.$} \sum\limits_{a}e^{i\alpha _{a}}\hat{Z}%
_{a\,i}-e^{-i(\alpha _{0}+\sum\limits_{a}\alpha _{a})}\hat{Z}_{i}%
\makebox[0.5em]{$\left.\rule{0pt}{1.1em}\right)$} \right] .
\end{equation}

In order to integrate the Maxwell Eqs.~(\ref{maxwellIntegrated}), we exploit
the same approach considered in Sec. ~\ref{sectionComposite}; the
flow-defining functions~$W_{i}$ and~$\chi _{i}$, as well as the
electromagnetic potential~$b^{\alpha }$, can again be represented as in~(\ref%
{WfasReIm}) and~(\ref{balmost}) in terms of symplectic vector~$T$, which for
the class under consideration has a slightly more complicated expression,
namely:
\begin{equation}
T^{\alpha }=-\frac{1}{4}\,\left( e^{-i\sum\limits_{a}\alpha _{a}}\bar{V}%
^{\alpha }-3e^{-i\alpha _{0}}V^{\alpha }+\sum\limits_{a}e^{-i\alpha _{a}}%
\bar{V}_{\overline{a}}^{\alpha }+e^{-i(\alpha _{0}+\sum\limits_{a}\alpha
_{a})}\sum_{a}e^{i\alpha _{a}}V_{{a}}^{\alpha }\,\right) .  \label{Talmost-1}
\end{equation}

When replacing the electromagnetic potential~$b$ expressed in terms of $T$ (%
\ref{Talmost-1}) into Eq.~(\ref{fAlmost}), this latter is turned into a
constrant:
\begin{equation}
\text{Im}\left[ e^{-i\alpha _{0}}\left( 3+e^{i(\alpha
_{0}+\sum\limits_{a}\alpha _{a})}\right) Z_{i}+\left( 1-e^{-i(\alpha
_{0}+\sum\limits_{a}\alpha _{a})}\right) \sum\limits_{a}e^{i\alpha
_{a}}Z_{a\,i}\right] =0.  \label{constrAlmost}
\end{equation}%
In other words, as for the composite non-BPS class treated in\ Sec. \ref%
{sectionComposite}, if one tries to deduce~$\chi _{i}$ from~(\ref{fAlmost})
recalling that (\textit{cfr.}~(\ref{maxwellIntegrated}) and~(\ref{Zhat}))
\begin{equation}
\hat{Z}_{i}=Z_{i}+\chi _{i}\langle b,V\rangle ,\qquad \hat{Z}%
_{a\,i}=Z_{a\,i}+\chi _{i}\langle b,V_{{a}}\rangle ,
\end{equation}%
then it can be checked that the dependence of~(\ref{fAlmost}) on~$\chi _{i}$
drops out, thus yielding the constraint~(\ref{constrAlmost}).

However, differently from the composite non-BPS class, in the almost BPS
class, the Maxwell equations turn out to be consistent \textit{only when two
further constraints are satisfied}, namely:%
\begin{equation}
\begin{array}{l}
\mathbf{I}:\alpha _{0}+\sum_{a}\alpha _{a}=\pi \qquad\mbox{or}\qquad \mathbf{%
II}:\alpha _{0}+\sum_{a}\alpha _{a}=2\arctan \left( e^{2U}B\right) ,%
\end{array}
\label{twocases}
\end{equation}%
thus giving rise, as in the $t^{3}$ model investigated in~\cite{Yeranyan-t^3}%
, to two distinct almost BPS branches\footnote{%
It is interesting to notice that each branch admits a single-centered limit
of different type: while branch $\mathbf{I}$ admits a particular BPS BH with
$i_{3}=0$ (\textit{cfr.} Sec. \ref{Branch-I}), branch $\mathbf{II}$ exhibits
a non-BPS BH (as single-centered limit of the particular set of embedded
composite non-BPS solutions discussed in Sec. \ref{Branch-II}). This is
consistent with the fact that two-centered almost BPS solutions are
characterized by a BPS and a non-BPS center~\cite%
{Bossard:2011kz,CFMY-Small-1}.}, which we will separately analyzed in the
next two Subsections.

The counting of the two ``flat'' directions along the (non-BPS, ``large'')
flows in the $stu$ model~\cite{Bellucci:2008sv,FM-2} is retrieved by
considering that in this class the four partial differential Eqs.~(\ref%
{solDPhasesAlmost}) for the four phases $\alpha _{0}$ and $\alpha _{a}$'s
are supplemented by the algebraic constraint~(\ref{constrAlmost}) and by the
further condition $\mathbf{I}$ or $\mathbf{II}$ of~(\ref{twocases}).

\subsection{\label{Branch-I}Branch $\mathbf{I}$}

In the case $\mathbf{I}$ of~(\ref{twocases}), the $D=3$ Cartesian rotation
vector reads
\begin{equation}
\chi _{i}=\partial _{i}B+e^{U}B\,\text{Re}\,\makebox[0.5em]{$\left[%
\rule{0pt}{1.1em}\right.$} \sum\limits_{a}e^{i\alpha
_{a}}Z_{a\,i}+3e^{-i\alpha _{0}}Z_{i}\makebox[0.5em]{$\left.%
\rule{0pt}{1.1em}\right]$} +e^{-U}\,\text{Im}\,\makebox[0.5em]{$\left[%
\rule{0pt}{1.1em}\right.$} \sum\limits_{a}e^{i\alpha
_{a}}Z_{a\,i}-e^{-i\alpha _{0}}Z_{i}\makebox[0.5em]{$\left.\rule{0pt}{1.1em}%
\right]$} ,  \label{chi-I}
\end{equation}%
and expressions for the flow-defining functions gets slightly simplified as
follows:
\begin{equation}
\begin{array}{l}
\displaystyle W_{i}=\text{Re}\makebox[0.5em]{$\left[\rule{0pt}{1.1em}%
\right.$} e^{-i\alpha _{0}}\hat{Z}_{i}\makebox[0.5em]{$\left.%
\rule{0pt}{1.1em}\right]$} , \\
\displaystyle \Pi _{i}^{a}=\frac{1}{2}\,e^{i\alpha _{0}}\hat{\overline{Z}}%
_{a\,i}-\frac{i}{2}\,e^{i(\alpha _{0}+\alpha _{a})}\,\text{Im}%
\makebox[0.5em]{$\left[\rule{0pt}{1.1em}\right.$} e^{-i\alpha _{0}}\hat{Z}%
_{i}\makebox[0.5em]{$\left.\rule{0pt}{1.1em}\right]$} +\frac{1}{2}%
\,e^{i(\alpha _{0}+\alpha _{a})}\,\text{Re}\makebox[0.5em]{$\left[%
\rule{0pt}{1.1em}\right.$} \sum\limits_{b}e^{i\alpha _{b}}\hat{Z}_{b\,i}%
\makebox[0.5em]{$\left.\rule{0pt}{1.1em}\right]$} .%
\end{array}%
\end{equation}

This branch contains BPS multi-centered solutions~\cite{Denef:2000nb} with a
particular constraint. This can be proved by simply setting $B=0$ in~(\ref%
{chi-I}), thus implying the functions~$W_{i}$ and~$\Pi _{i}^{a}$ to turn
into their BPS counterparts~(\ref{WPiBPS}), whit rotation~$\chi _{i}$
acquiring the following form:
\begin{equation}
\chi _{i}=e^{-U}\text{Im}\,\makebox[0.5em]{$\left[\rule{0pt}{1.1em}\right.$}
e^{i\sum\limits_{a}\alpha _{a}}Z_{i}+\sum_{a}e^{i\alpha _{a}}Z_{i\,a}%
\makebox[0.5em]{$\left.\rule{0pt}{1.1em}\right]$} .  \label{almostBPS_rot}
\end{equation}%
Independently on the value of the~$B$-field, the constraint~(\ref%
{constrAlmost}) takes the form
\begin{equation}
\text{Im}\,\makebox[0.5em]{$\left[\rule{0pt}{1.1em}\right.$}
e^{i\sum\limits_{a}\alpha _{a}}Z_{i}-\sum_{a}e^{i\alpha _{a}}Z_{i\,a}%
\makebox[0.5em]{$\left.\rule{0pt}{1.1em}\right]$} =0.  \label{addConstr1}
\end{equation}%
Thus,~(\ref{almostBPS_rot}) and~(\ref{addConstr1}) yield
\begin{equation}
\chi _{i}=-2\,e^{-U}\,\text{Im}\makebox[0.5em]{$\left[\rule{0pt}{1em}%
\right.$} e^{-i\alpha _{0}}Z_{i}\makebox[0.5em]{$\left.\rule{0pt}{1em}%
\right]$} =2\,e^{-U}\,\text{Im}\,\makebox[0.5em]{$\left[\rule{0pt}{1.1em}%
\right.$} \sum_{a}e^{i\alpha _{a}}Z_{ai}\makebox[0.5em]{$\left.%
\rule{0pt}{1.1em}\right]$} ,
\end{equation}%
which is exactly the expression of the $D=3$ rotation pertaining to the BPS
multi-centered class~\cite{Denef:2000nb} subjected to the additional
constraint~(\ref{addConstr1}).

The actual meaning of the constraint~(\ref{addConstr1}) has yet to be
clarified; nevertheless, we can here make some observations. Considering the
corresponding single-centered case, and thus setting both $B$ and $\chi _{i}$
to zero, this constraint splits in two parts, namely:
\begin{equation}
\text{Im}\makebox[0.5em]{$\left[\rule{0pt}{1.1em}\right.$}\sum_{a}e^{i\alpha
_{a}}Z_{a\,i}\makebox[0.5em]{$\left.\rule{0pt}{1.1em}\right]$} =0\qquad %
\mbox{and}\qquad \text{Im}\makebox[0.5em]{$\left[\rule{0pt}{1.1em}\right.$}
e^{i\sum\limits_{a}\alpha _{a}}Z_{i}\makebox[0.5em]{$\left.\rule{0pt}{1.1em}%
\right]$} =0.
\end{equation}%
In this case, \textit{at least} a particular solution
\begin{equation}
\alpha _{a}=-\arg Z_{a} ,\qquad \arg Z=\sum_{a}\arg Z_{a}
\end{equation}%
of the constraint $\mathbf{I}$ of~(\ref{twocases}) can be shown to be
consistent with the first order equations and, as a consequence, the duality
invariant~$i_{3}$~\cite{CFMZ1,Ceresole:2009iy} vanishes. In this sense, the
constraint~(\ref{addConstr1}) can be regarded as the generalization, within
the BPS multi-centered flow embedded into the branch $\mathbf{I}$ of the
almost BPS class, of the constraint $i_{3}=0$.

At the moment, it is not known whether this branch contains other consistent
multi-centered solutions different from the aforementioned constrained BPS
solutions; we leave this issue for further investigation.

\subsection{\label{Branch-II}Branch $\mathbf{II}$}

In the case $\mathbf{II}$ of~(\ref{twocases}), the $D=3$ Cartesian rotation
vector reads
\begin{equation}
\chi _{i}=\partial _{i}B-Be^{U}\,\text{Re}\makebox[0.5em]{$\left[%
\rule{0pt}{1.1em}\right.$} 3\,e^{i\sum\limits_{a}\alpha
_{a}}Z_{i}-\sum_{a}e^{i\alpha _{a}}Z_{a\,i}\makebox[0.5em]{$\left.%
\rule{0pt}{1.1em}\right]$} +e^{-U}\text{Im}\makebox[0.5em]{$\left[%
\rule{0pt}{1.1em}\right.$} 3\,e^{i\sum\limits_{a}\alpha
_{a}}Z_{i}-\sum_{a}e^{i\alpha _{a}}Z_{a\,i}\makebox[0.5em]{$\left.%
\rule{0pt}{1.1em}\right]$} ,  \label{falmost2}
\end{equation}%
and the constraint~(\ref{constrAlmost}) acquires the following form:
\begin{equation}
\begin{array}{l}
\displaystyle 2\,\text{Im}\makebox[0.5em]{$\left[\rule{0pt}{1.1em}\right.$}
e^{i\sum\limits_{a}\alpha _{a}}Z_{i}\makebox[0.5em]{$\left.\rule{0pt}{1.1em}%
\right]$} -B\,e^{2U}\text{Re}\makebox[0.5em]{$\left[\rule{0pt}{1.1em}%
\right.$} 3\,e^{i\sum\limits_{a}\alpha _{a}}Z_{i}-\sum_{a}e^{i\alpha
_{a}}Z_{a\,i}\makebox[0.5em]{$\left.\rule{0pt}{1.1em}\right]$} \\
\displaystyle\phantom{2\,\text{Im}\LSB{0.5em}{1.1em}
e^{i\sum\limits_{a}\alpha _{a}}Z_{i}\RSB{0.5em}{1.1em}} -B^{2}e^{4U}\text{Im}%
\makebox[0.5em]{$\left[\rule{0pt}{1.1em}\right.$} e^{i\sum\limits_{a}\alpha
_{a}}Z_{i}-\sum_{a}e^{i\alpha _{a}}Z_{a\,i}\makebox[0.5em]{$\left.%
\rule{0pt}{1.1em}\right]$} =0.%
\end{array}
\label{constrAlmost2}
\end{equation}%
Here we refrain from writing down the explicit expressions for the
flow-defining functions $W_{i}$ and $\Pi _{i}^{a}$, resulting from plugging
Eq.~(\ref{twocases}) (case $\mathbf{II}$) into~(\ref{defAlmost}), because in
this branch their form is not very illuminating and rather cumbersome.

It is worth remarking that, among all possible solutions to Eqs.~(\ref%
{falmost2})-(\ref{constrAlmost2}), there is a particular one that coincides
with a particular solution of the composite non-BPS treated in\ Sec \ref%
{sectionAlmost}; indeed, if the $B$-field is set to a constant and the
rotation~$\chi _{i}$ vanishes, then it can be checked that the set of
equations~(\ref{falmost2})-(\ref{constrAlmost2}) turns into the set~\ of
Eqs.~(\ref{constrComp}) and~(\ref{deffcomposit2}), which describes the
composite non-BPS class.

\paragraph{An Example with Fixed ``Flat'' Directions}

An interesting example is provided by the choice of all phases~$\alpha _{a}$%
's to be equal. The consistency of Eqs.~(\ref{solDPhasesAlmost}) within the
case $\mathbf{II}$ of~(\ref{twocases}) requires then that
\begin{equation}
\alpha _{1}=\alpha _{2}=\alpha _{3}=-\frac{\pi }{2}+\arctan Be^{2U},\quad
\alpha _{0}=\frac{3\pi }{2}-\arctan Be^{2U}.
\end{equation}%
This latter exactly falls into the class of solutions \cite%
{Goldstein-Katmadas,Bossard:2011kz}
\begin{equation}
e^{-4U}=4H^{0}\mathcal{Z}_{1}\mathcal{Z}_{2}\mathcal{Z}_{3}-B^{2},\quad
z^{a}=K^{a}-\frac{B-ie^{2U}}{2H^{0}\mathcal{Z}_{a}},
\end{equation}%
where $H^{0}$ is harmonic, and the functions~$\mathcal{Z}_{a}$ and~$K^{a}$
satisfy the following equations:
\begin{equation}
\partial _{i}\mathcal{Z}_{a}=H_{i\,a}-|\epsilon _{abc}|\,\left( H_{i}^{b}-%
\frac{1}{2}K^{b}\partial _{i}H^{0}\right) K^{c},\qquad \partial _{i}K_{a}=-%
\frac{1}{H^{0}}\left( H_{i}^{a}-K^{a}\partial _{i}H^{0}\right) ,
\end{equation}%
where the latter equation implies $K^{a}$ to be harmonic. In this case, the $%
D=3$ Cartesian rotation vector $\chi _{i}$ can be rewritten as%
\begin{equation}
\chi _{i}=\partial _{i}B-2H^{0}\partial _{i}K^{a}\mathcal{Z}_{a},
\end{equation}%
and the constraint (\ref{constrAlmost}) reads
\begin{equation}
H_{i\,0}+H_{i\,a}K^{a}+\frac{1}{2}|\epsilon _{abc}|\partial
_{i}K^{a}K^{b}K^{c}H^{0}-\partial _{i}K^{a}\mathcal{Z}_{a}-2K^{1}K^{2}K^{3}%
\partial _{i}H^{0}=0.
\end{equation}

This solution can generate other solutions by applying the $U$-duality and
Ehlers symmetry transformations, as considered in~\cite{Bossard:2011kz}. We
leave to further future investigation the issue of existence of general
multi-centered solutions of this class; for instance, in the two-centered
case, such general solutions should exhibit all the duality-invariant
polynomials of the \textquotedblleft minimal degree\textquotedblright\
complete basis~\cite{FMOSY-1} as \textit{independent}.

\section{\label{Conclusion}Conclusion}

In the present investigation, we developed the most general first order
formalism for multi-centered and/or under-rotating extremal BH solutions
with flat three-dimensional base-space in the $stu$ model~\cite%
{STU,Bellucci:2008sv} of $N=2$, $D=4$ ungauged Maxwell-Einstein
supergravity. As mentioned, this is a universal sector of all $N\geqslant 2$%
-extended supergravity theories with symmetric (vector multiplets') scalar
manifold, and which admit an uplift to $D=5$ dimensions.

Our procedure, which generalizes the one exploited for the simpler $t^{3}$
model in~\cite{Yeranyan-t^3}, sets the non-supersymmetric (composite non-BPS
\cite{Bossard:2011kz} and almost BPS~\cite{Goldstein-Katmadas,Bossard:2011kz}%
) classes of multi-centered solutions on the same footing of the well known
BPS class~\cite{Denef:2000nb,Denef-2}, thus allowing for a unified framework
for the study of the flow dynamics of scalar fields.

We developed a tree-dimensional Cartesian formalism in which the effective
BH potential~\cite{Ferrara:1997tw}, as well as the scalar-dependent central
and matter charges and the first order ``fake'' superpotential \cite{CD-1},
are generalized in a not necessarily axisymmetric framework (\textit{cfr.}~(%
\ref{V-Cart}), (\ref{Zhat}),~(\ref{Ress-1}) and~(\ref{alg-constr-1})), which
thus allows to handle also (under-)rotating and multi-centered BH solutions.

Then, by extending some spatially asymptotical expressions all along the
flow, we derived systems of partial differential equations for the $\alpha $%
-phases describing each class of multi-centered systems. The corresponding
counting, supplemented by an algebraic constraint (and by a further
condition in the almost-BPS class) is consistent with the existence of two
``flat'' directions along the non-rotating sigle-centered $stu $ non-BPS
flow~\cite{Bellucci:2008sv,FM-2}.

The consistency of the systems of partial differential equations on which
the first order formalism is based has then been checked also by retrieving
known solutions, in the single-centered limit (such as the Rasheed-Larsen
non-BPS Kaluza-Klein BH~\cite{Rasheed:1995zv, Larsen:1999pp} in the
composite non-BPS class) as well as in the multi-centered case (various
solutions from~\cite{Denef-2,Bossard:2011kz,Bossard-Octonionic} have been
obtained as particular solutions).

Consistent with its nilpotent orbit characterization~\cite{Bossard:2011kz}
as well as with the results obtained in~\cite{Yeranyan-t^3} for the $t^{3}$
model, the almost BPS class~\cite{Goldstein-Katmadas,Bossard:2011kz}
exhibits the most involved structure: indeed, it is described by four
independent phases $\alpha _{0}$ and $\alpha _{a}$ ($a=1,2,3$), and in this
class the consistency of Maxwell equations further imposes the constraint (%
\ref{twocases}), whose two-fold nature gives rise to two sub-branches. Such
a split can be traced back to the possibility to have both BPS and non-BPS
BH centers in this class, which thus stands on a different footing with
respect to the BPS and composite non-BPS classes, in which the BH centers
are of the same type.

As mentioned in Secs.~\ref{sectionComposite} and~\ref{sectionAlmost}, we
leave for future investigation the issue of existence of completely general
solutions (with all electric-magnetic charges switched on in each BH center)
in the composite non-BPS \textit{and/or} almost BPS classes. In the
two-centered case, such general solutions would exhibit all independent
duality-invariant polynomials of the complete \textquotedblleft minimal
degree\textquotedblright\ basis~\cite{FMOSY-1,CFMY-Small-1,FMY-CV-1}.
Moreover, it would be interesting to analyze the issue of the fixed distance
among the BH centers in the non-supersymmetric systems~\cite%
{Bena,Bossard:2011kz,Bossard-Octonionic}, also in relation to the existence
of walls of marginal stability.

It is here worth pointing out that we considered only multi-centered
configurations with ``large'' BH centers, namely with a well-defined
near-horizon geometry (and thus, scalar dynamics) already at the Einstein
(two-derivatives) level; at the moment, it is not clear whether some
solutions with ``small'' BH centers (cfr. \textit{e.g.} \cite{GLS-2,CS-1})
can be obtained in this framework (for an analysis in the two-centered case
at the level of duality invariants, see~\cite{CFMY-Small-1}).

The possible extension of our approach to multi-centered systems with
\textit{non-flat} three-dimensional base-space is of utmost interest, as
well; within this framework, one should recover solutions found \textit{e.g.}
in~\cite{Bena}. It would also be interesting to consider generalizations to
\textit{over-rotating} (single- and multi- centered) BH solutions (for
recent advances on first order formalism, see \textit{e.g.} \cite{T}).

Our approach could also be applied to the multi-centered interacting non-BPS
solutions of maximal $N=8$, $D=4$ supergravity recently considered \textit{%
e.g.} in~\cite{Bossard-Octonionic}, and the issue of truncability to the
universal $stu$ sector is also worth being investigated in greater detail.

\section*{Acknowledgements}

We would like to thank G. Bossard, A. Ceresole and G. Dall'Agata for useful
discussions and correspondence.

The work of S.F. and A. Y. has been supported by the ERC Advanced Grant no.
226455, Supersymmetry, Quantum Gravity and Gauge Fields (SUPERFIELDS).

\appendix

\section{\label{notations}Notations and Useful Formul\ae }

We here recall some useful notations and formul\ae , used throughout the
present investigation.

In a \textit{generic} special K\"{a}hler geometry, the symmetric real matrix~%
$M$, which firstly occurred in our treatment in Eq.~(\ref{d3act3}), is
constructed from the coupling matrices~$\mu _{\Lambda \Sigma }$ and~$\nu
_{\Lambda \Sigma }$ as follows~\cite{CDF-rev,FK}
\begin{equation*}
M_{\alpha \beta }=\left(
\begin{array}{cc}
\mu _{\Lambda \Sigma }+\nu _{\Lambda \Lambda ^{\prime }}\mu ^{\Lambda
^{\prime }\Sigma ^{\prime }}\nu _{\Sigma ^{\prime }\Sigma } & \nu _{\Lambda
\Lambda ^{\prime }}\mu ^{\Lambda ^{\prime }\Sigma } \\
\mu ^{\Lambda \Lambda ^{\prime }}\nu _{\Lambda ^{\prime }\Sigma } & \mu
^{\Lambda \Sigma }%
\end{array}%
\right) ,
\end{equation*}%
%
%
%
%
%
and it is symplectic:
\begin{equation*}
\Omega ^{\alpha \beta }M_{\beta \gamma }=-M^{\alpha \beta }\Omega _{\beta
\gamma }
\end{equation*}%
%
%
%
%
%
with respect to the skew-symmetric symplectic metric
\begin{equation*}
\Omega _{\alpha \beta }=\left(
\begin{array}{cc}
0 & -\delta _{\Lambda }^{\Sigma } \\
\delta _{\Sigma }^{\Lambda } & 0%
\end{array}%
\right) ,\qquad \Omega ^{\alpha \beta }=\left(
\begin{array}{cc}
0 & \delta _{\Sigma }^{\Lambda } \\
-\delta _{\Lambda }^{\Sigma } & 0%
\end{array}%
\right) ,  \label{sympl-metric}
\end{equation*}%
%
%
%
%
%
which allows one to define the \textit{symplectic product} of two vectors
as:
\begin{equation*}
\langle A,B\rangle \equiv A^{\alpha }\Omega _{\alpha \beta }B^{\beta
}=A_{\Lambda }B^{\Lambda }-A^{\Lambda }B_{\Lambda }.
\end{equation*}%
%
%
%
%
%

The inverse of the coupling matrix can be easily calculated to read
\begin{equation*}
M^{\alpha \beta }=\left(
\begin{array}{cc}
\mu ^{\Lambda \Sigma } & -\mu ^{\Lambda \Lambda ^{\prime }}\nu _{\Lambda
^{\prime }\Sigma } \\
-\nu _{\Lambda \Lambda ^{\prime }}\mu ^{\Lambda ^{\prime }\Sigma } & \mu
_{\Lambda \Sigma }+\nu _{\Lambda \Lambda ^{\prime }}\mu ^{\Lambda ^{\prime
}\Sigma ^{\prime }}\nu _{\Sigma ^{\prime }\Sigma }%
\end{array}%
\right).
\end{equation*}%
%
%
%
%
%
The special geometry sections are normalized as%
\begin{equation*}
\langle V,\bar{V}\rangle =-i,\quad \langle V,D_{a}V\rangle =0,\quad \langle
D_{a}V,\bar{D}_{\bar{a}}\bar{V}\rangle =iG_{a\bar{a}}.
\end{equation*}%
%
%
%
%
%
The coupling matrix~$M$ satisfy the following identity
\begin{equation*}
\frac{1}{2}(M_{\alpha \beta }-i\Omega _{\alpha \beta })=-(\Omega _{\alpha
\alpha ^{\prime }}\bar{V}^{\alpha ^{\prime }})(\Omega _{\beta \beta ^{\prime
}}V^{\beta ^{\prime }})-(\Omega _{\alpha \alpha ^{\prime }}D_{a}V^{\alpha
})G^{a\bar{a}}(\Omega _{\beta \beta ^{\prime }}\bar{D}_{\bar{a}}\bar{V}%
^{\beta ^{\prime }}).
\end{equation*}
The $C$-tensor of special K\"{a}hler geometry and the metric of the vector
multiplets' scalar manifold enter the basic relations
\begin{equation*}
D_{a}D_{b}V=iC_{abc}G^{c\bar{c}}\bar{D}_{\bar{c}}\bar{V},\quad D_{a}\bar{D}_{%
\bar{b}}\bar{V}=G_{a\bar{b}}\bar{V}.
\end{equation*}%
%
%
%
%
%
In the $stu$ model
\begin{equation*}
F=D_{abc}z^{a}z^{b}z^{c} = s t u,
\end{equation*}
the Vielbein and its inverse (we maintain here the underlining of flat
scalar indices)
\begin{equation*}
E_{a}{}^{\underline{{a}}}=\text{diag}\rule{0em}{1em}\left( \frac{1}{s-\bar{s}%
}\,,\frac{1}{t-\bar{t}}\,,\frac{1}{u-\bar{u}}\right) \rule{0em}{1em},\qquad
E_{{\underline{a}}}{}^{a}=\text{diag}\left( s-\bar{s},t-\bar{t},u-\bar{u}%
\right)
\end{equation*}%
%
%
%
%
%
satisfy the usual definition
\begin{equation*}
G_{a\bar{a}}=E_{a}{}^{{\underline{a}}}\,\delta_{{\underline{a}}\,\bar{%
\underline{a}}}{\bar{E}}_{\bar{a}}{}^{\bar{\underline{a}}},\qquad \delta_{{%
\underline{a}}\,\bar{\underline{a}}}=\text{diag}(1,1,1).
\end{equation*}%
%
%
%
%
%

\end{document}